\documentclass[aps,prb,floatfix,twocolumn,citeautoscript,showpacs,notitlepage]{revtex4-1}

\usepackage[english]{babel}
\addto\captionsenglish{}
\usepackage{amsmath,amssymb,amsthm,amsfonts}
\usepackage{graphicx}
\usepackage{dcolumn}
\usepackage{bm}
\usepackage{tabularx}
\usepackage{xspace}
\usepackage{txfonts}
\usepackage{ulem}
\usepackage[caption=false]{subfig}
\usepackage[hidelinks,colorlinks,allcolors=blue]{hyperref}

\newcommand{\fig}[1]{Fig.~\ref{#1}}
\newcommand{\app}[1]{Appendix~\ref{#1}}
\newcommand{\Sec}[1]{Sec.~\ref{#1}}
\newcommand{\eq}[1]{Eq.~\eqref{#1}}
\newcommand{\rref}[1]{Ref.~[\onlinecite{#1}]}
\newcommand{\rrefs}[1]{Refs.~[\onlinecite{#1}]}

\renewcommand{\ol}{\overline}
 
\newcommand{\mc}[1]{\mathcal{#1}}
\renewcommand{\d}{\mathrm{d}}
\newcommand{\ii}{\mathrm{i}}
\newcommand{\e}{\mathrm{e}}
\renewcommand{\H}{\mc{H}}
\renewcommand{\S}{\Omega}
\newcommand{\E}{\mc{E}}
\newcommand{\I}{\mc{I}}
\newcommand{\D}{\mc{D}}

\newcommand{\bR}{\textbf{R}}
\newcommand{\bk}{\textbf{k}}
\newcommand{\bQ}{\textbf{Q}}
\newcommand{\avg}[1]{\langle #1 \rangle}
\renewcommand{\dag}{\dagger}
\newcommand{\nodag}{{\vphantom\dag}}
\newcommand{\transp}[1]{#1^\mathrm{T}}
\newcommand{\cn}{c^{\nodag}}
\newcommand{\cd}{c^\dag}
\newcommand{\fn}{f^{\nodag}}
\newcommand{\fd}{f^\dag}
\newcommand{\zn}{z^{\nodag}}
\newcommand{\zd}{z^\dag}
\newcommand{\up}{\uparrow}
\newcommand{\down}{\downarrow}
\newcommand{\bmoy}{\ol{\beta}}
\newcommand{\mueff}{\tilde{\mu}}

\newcommand{\ed}{e^\dag}
\newcommand{\en}{e^\nodag}
\newcommand{\dd}{d^\dag}
\newcommand{\dn}{d^\nodag}
\newcommand{\pd}{p^\dag}
\newcommand{\pn}{p^{\nodag}}
\newcommand{\bn}{\beta^{\nodag}}
\newcommand{\an}{\alpha^{\nodag}}
\newcommand{\s}{\sigma}
\newcommand{\m}{\varepsilon}
\DeclareMathOperator{\re}{Re}
\renewcommand{\Re}[1]{\re\!\left( #1 \right)}
\DeclareMathOperator{\im}{Im}
\renewcommand{\Im}[1]{\im\!\left( #1 \right)}
\newcommand{\sump}[1]{\sum_{#1}{\vphantom{\sum}}'}

\def\subinrm#1{\sb{\rm#1}}
{\catcode`\_=13 \global\let_=\subinrm}
\mathcode`_="8000
\def\upsubscripts{\catcode`\_=12 } 
\upsubscripts

\def\supinrm#1{^{\rm#1}}
{\catcode`\^=13 \global\let^=\supinrm}
\mathcode`^="8000
\def\upsupscripts{\catcode`\^=12 } 
\upsupscripts

\begin{document}

\title{Spin and charge modulations of a half-filled extended Hubbard model}

\newcommand{\affcrismat}{\affiliation{
Normandie Universit\'e, ENSICAEN, UNICAEN, CNRS, CRISMAT, 14000 Caen, France}}

\author{Lo\"ic~Philoxene}

\author{Vu Hung~Dao}
\email{vu-hung.dao@ensicaen.fr}

\author{Raymond~Fr\'esard}
\affcrismat

\date{\today}

\begin{abstract}
We introduce and analyze an extended Hubbard model, in which intersite
Coulomb interaction as well as a staggered local potential (SLP) are considered,
on the square lattice at half band filling, in the thermodynamic limit. 
Using both Hartree-Fock approximation and Kotliar and Ruckenstein slave boson formalism, we
show that the model harbors charge order (CO) as well as joint spin and charge
modulations (SCO) at finite values of the SLP, while the spin density
wave (SDW) is stabilized for vanishing SLP, only. We determine their phase
boundaries and the variations of the order parameters in dependence on the
SLP, as well as on the on-site and nearest-neighbor
interactions. Domains of coexistence of CO and SCO phases, suitable for 
resistive switching experiments, are unraveled. We show that the novel SCO
systematically turns into the more conventional SDW phase when the
zero-SLP limit is taken.  We also discuss the nature of the different
phase transitions, both at zero and finite temperature. In the former case, 
no continuous CO to SDW (or SCO) phase transition occurs. In contrast, a
paramagnetic phase (PM), which is accompanied with continuous phase transitions
towards both spin or charge ordered phases, sets in at finite
temperature. A good quantitative agreement with numerical simulations is
demonstrated, and a comparison between the two used approaches is performed.
\end{abstract}

\maketitle


\section{Introduction}\label{sec:intro}

Electronic phases displaying spin and charge modulations are currently
undergoing intense scrutiny \cite{wu2021,campi2022}. Following the seminal works by Tranquada et
al. on doped nickelates\cite{tranquada1994,sachan1995}, they are firstly found
in doped Mott insulators. They received even stronger interest when related
order was evidenced in the superconducting Sr-doped La$_2$CuO$_4$ cuprates
(LSCO) \cite{tranquada1995}. Furthermore, striped states have been exhibited
in other series of oxides as well which, most notably, include layered
cobaltates \cite{cwik2009} and layered manganites \cite{ulbrich2012}, as
reviewed in, \textit{e.~g.}, \rref{ulbrich2012}. 

A diversity of stripe orders have been reported. Indeed, the wave-vectors
characterizing the modulations may either lie along the diagonal of the
Brillouin zone (BZ), in which case the stripe is coined diagonal, or along the
side of the BZ, and the stripe is said to be vertical. The stripes observed in
nickelates are diagonal and are systematically found to be insulating
\cite{cheong1994,huecker2007}. This holds true for layered cobaltates and 
layered manganites as well. Cuprates, in contrast, present both
diagonal and vertical stripes, with the former being insulating and the latter
metallic, if not superconducting\cite{huecker2011}. Explaining the insulating
nature of La$_{2-x}$Sr$_x$NiO$_4$ nickelates by Density Functional Theory,
including optimized lattice distortion
\cite{yamamoto2007,schwingen2008,schwingen2009}, or by means of model
calculations \cite{raczkowski2006_a}, turned out to be
challenging. Conversely, for cuprates, a systematics of insulating
diagonal filled (with nearly 1 hole per domain wall) and metallic half-filled
(with nearly 0.5 hole per domain wall) stripes could be established within
model calculations \cite{raczkowski2006_b}. More recent numerical calculations applying a variety of
approaches tackled the issue of d-wave superconductivity in the ground state
of the two-dimensional Hubbard Model. As of today, it appears fair to say that
it could not unambiguously be established. Actually, stripe order has been
found to be very close, if not lower, in energy to its superconducting
counterpart
\cite{himeda2002,ido2018,white2009,huang2018,corboz2014,ponsioen2019,li2021}
--- coexistence of both phases has been highlighted too
\cite{leprevost2015,jiang2020} albeit not necessarily in the ground state
\cite{zheng2017}. 

So far we briefly presented spin and charge modulated phases of the doped
two-dimensional Hubbard model, but would they persist in the half band filling
limit? No positive answer to this question resulted from the study of the
two-dimensional Hubbard model which, for repulsive on-site interaction $U>0$,
only harbors antiferromagnetism. This could be presumed considering the
following qualitative picture of the stripe modulations. The spatial periods
of the charge order in the striped phase are bounded by holes, or fractions of
holes. A higher hole concentration thus eventually leads to a shorter spatial
period of the stripe ordering, and conversely. By taking the limit where the
hole doping tends to zero, the period of the stripe order diverges such that
we are left with the usual antiferromagnetic Mott insulating phase at half
band-filling. This heuristic argument supports the assumption that the
mechanisms underlying spin-and-charge modulations at half filling should
differ from those underlying the striped phases at finite
doping. Additionally, evidence of the existence of a spin-and-charge ordered
phase in the half-filled Hubbard model remains elusive \cite{vanDongen1994a,
vanDongen1994b,wolff1983,deeg1993,hirsch1984,segpunta2002,zhang1989,
terletska2017,paki2019}. As a consequence to this, in the present work, a 
Hubbard model extended with a nearest-neighbor interaction term of intensity 
$V$ as well as a spatial modulation of the energy levels $\m_i$ in order to 
straightforwardly induce charge order in the antiferromagnetic phase is 
considered, and phases bridging between pure magnetic order and pure charge 
order are sought for. 

The rich landscape of competing electronic phases displayed by metal oxides is not
their only point of interest. When the transition between different states is of
the first order, it can be harnessed for application in digital electronics. Since the
functionnal materials can keep their valuable properties down to the nanoscale,
they promise to offer a superior alternative to conventional semiconductor
components~\cite{takagi2010,coll2019}.
In particular, resistive switching in Mott systems is the subject of intense investigations, 
since it could enable a variety of novel functions, such as resistive RAM for data 
storage~\cite{janod2015,wang2019}, optoelectronics~\cite{liao2018,coll2019}, or 
neuromorphic computing~\cite{delvalle2018,delvalle2019,bauers2021}. Their switch 
between resistivity values which may differ by several orders of magnitude,
can be experimentally triggered by varying different control parameters, such
as chemical doping, strain, temperature, hydrostatic pressure, electric field,
current, and illumination. The microscopic mechanisms that drive the phase
transition are, however, still 
debated~\cite{diener2018,kalcheim2020,babich2022,adda2022}. Yet the diversity
of stimuli indicates that several are likely at play depending on the
material, and it highlights the versatility of the phenomenon as a tool for
future technologies. 

In this context, we analyze our model by performing both Hartree-Fock (HF) and Kotliar-Ruckenstein 
slave boson (KRSB) slave boson calculations. 
While the former method mostly yields qualitative understanding, 
the latter better incorporates correlation effects.

This paper is organized as follows: in \Sec{sec:model} the considered
Hamiltonian, the HF approximation as well as the saddle-point approximation to the 
KRSB representation used in this study are presented. 
In \Sec{sec:res_hf} within the HF approximation, 
and in \Sec{sec:res_krsb} within the KRSB representation, the zero-temperature phase diagram of 
the half-filled model on the square lattice is unraveled for different values of the splitting 
$\m$ and a novel spin-and-charge ordered phase is presented. In \Sec{sec:tuv_krsb}, within the 
KRSB representation, the $\m = 0$ limit is taken in order to recover the $t$--$U$--$V$ model 
and the phase diagram is presented for zero and finite temperature. 
We then compare, in \Sec{sec:comp}, the results obtained within both formalisms in order to assess for 
the relevance of electronic correlations in the presented phases.
Finally, \Sec{sec:concl} 
presents conclusions and a short outlook. \app{app:hf} details the self-consistent 
field equations to be solved in the HF approximation. \app{app:saddle_point} gives more details 
about the setup of the different saddle-points, together with the derivation of the saddle-point 
equations.

\section{Model and methods}\label{sec:model}

\subsection{Extended Hubbard model}

Initially introduced nearly simultaneously by Hubbard~\cite{hubbard1963}, 
Kanamori~\cite{kanamori1963} and Gutzwiller~\cite{gutzwiller1963} as a model for 
electrons in transition-metal oxides, the now-called Hubbard model is the 
archetype of correlated electrons model. Its Hamiltonian embodies the competition
between kinetic energy and Coulomb interactions in its simplest form. It reads
\begin{equation}\label{model:HH}
    \H_{Hub} = \H_0 + \H_U ,
\end{equation}
with the one-body part, in the grand canonical ensemble,
\begin{equation}
    \H_0 = \sum_{i,j,\s} t^{\phantom{0}}_{ij} \cd_{i,\s} \cn_{j,\s} -\mu \sum_{i,\s} \cd_{i,\s} \cn_{i,\s} ,
\end{equation}
and the local interaction term
\begin{equation}
    \H_U = U \sum_i \cd_{i,\up} \cd_{i,\down} \cn_{i,\down} \cn_{i,\up} ,
\end{equation}
where $\cd_{i,\s}$ ($\cn_{i,\s}$) creates (annihilates) an electron with spin $\s \in \{ \up,\down \}$ at the lattice site i, $n_{i,\s}$ is the usual electron number operator ($n_i=n_{i,\up}+n_{i,\down}$), $t^{\phantom{0}}_{ij}$ is the hopping amplitude between lattice sites i and j, $\mu$ is the chemical potential controlling the band filling and $U$ is the on-site Coulomb interaction strength. Nevertheless, Hubbard himself pointed out the major drawback of his model when it comes to faithfully describing the microscopic processes happening in real systems, namely the over-simplified treatment of the Coulomb interaction \cite{hubbard1964}. In cuprates, for example, it appears that the screening of the Coulomb interaction between electrons is not perfect and thus yields inter-atomic contributions. As a response to this, longer-ranged Coulomb interactions may be incorporated in the Hamiltonian, yielding the additional term
\begin{equation}\label{model:HV}
    \H_V = \frac{1}{2} \sum_{\substack{\rm i,j\\ \s,\s'}} V_{ij} \cd_{i,\s} \cd_{j,\s'} \cn_{j,\s'} \cn_{i,\s} ,
\end{equation}
where $V_{ij}$ is the Coulomb coupling between sites i and j. The $1/2$
factor accounts for the double-counting of the $\rm (i,j)\equiv(j,i)$ pairs in the
sum. Henceforth, we assume the screening of the long-ranged Coulomb
interaction to be efficient, and we choose $V_{ij}$ such that $V_{ij}=V$ if i and 
j are nearest-neighbors and $V_{ij}=0$ otherwise. Approximating the Coulomb 
interaction in such a way yields the often-called extended or $t$--$U$--$V$ Hubbard 
model 
\begin{equation} \label{model:Htuv}
    \H_{t-U-V}=\H_{Hub}+\H_V.
\end{equation}

Turning next to our search for a minimal model of spin-and-charge modulated phases 
at half-filling, we propose to further extend the $t$--$U$--$V$ model by also 
considering an staggered local potential (SLP) of the form
\begin{equation}
    \H_{\m} = \sum_{i,\s} \m_i \cd_{i,\s} \cn_{i,\s} .
\end{equation}
This potential either arises from an inhomogeneous crystal field inside a material 
or an alternating intensity of the confinement potential in the context of cold 
fermions on an optical lattice.
Models of strong disorder, where $\m_i$ is randomly distributed across the lattice 
such that each site has a different electrostatic environment, has been considered 
in the context of many-body localization \cite{prelovsek2008,prelovsek2021}. In 
the present case however, the focus is made on minimal extensions to the Hubbard 
model. We thus consider the simple form
\begin{subequations}
\begin{align}
    &\m_i = -\m \quad\text{if}\quad \rm i \in A , \\
    &\m_i = +\m \quad\text{if}\quad \rm i \in B ,
\end{align}
\end{subequations}
where A and B are the two sublattices of a bipartite lattice. In the
context of LSCO superconductors, alternating La and, \textit{e.~g.}, Pr chains,
could produce such a contribution to the crystal field, as the electronic cloud of the La 
cations is more extended than the one of the Pr cations. 
This results in a staggered $2\m$ energy distribution
between adjacent lattice sites. This finally leads us to the Hamiltonian 
\begin{equation} \label{model:Htot}
    \H_{ext} = \H_{Hub} + \H_V + \H_{\m} ,
\end{equation}
which will be investigated in \Sec{sec:res_hf} and in \Sec{sec:res_krsb}.

Throughout the present study, we work in the half-filled subspace of the square 
lattice, such that
\begin{equation}
    \sum_{i,\s} \langle n_{i,\s} \rangle=  N_L ,
\end{equation}
where $N_L$ is the total number of lattice sites and we set the lattice parameter 
$a=1$ thereby fixing the length scale.

\subsection{HF approximation}

Throughout this work, we use the HF approximation as a benchmark for the weak coupling regime and compare its results in the intermediate-to-strong coupling regime with the KRSB results including correlations. In this framework, the interaction terms in \eq{model:Htuv} are approximated by
\begin{equation}
    \H_{U} \approx U \sum_{i,\s} \left( \avg{n_{i,-\s}} \cd_{i,\s} \cn_{i,\s} - \frac{1}{2} \avg{n_{i,\up}} \avg{n_{i,\down}} \right),
\end{equation}
and
\begin{align}
    \H_{V} \approx V \sum_{\avg{ij},\s} & \left[ \sum_{\s'} \left( \avg{n_{j,\s'}} \cd_{i,\s} \cn_{i,\s} - \frac{1}{2} \avg{n_{i,\s}} \avg{n_{j,\s'}} \right) \right. \notag\\ 
    &\left. - \frac{1}{2}\avg{b_{ij,\s}} \left( \cd_{i,\s} \cn_{j,\s} + \rm{h.c.} \right) + \frac{1}{2} \avg{b_{ij,\s}}^2 \right] ,
\end{align}
where
\begin{subequations}
\begin{align}
    \avg{n_{i,\s}} = \avg{\cd_{i,\s}\cn_{i,\s}} , \\
    \avg{n_i} = \avg{n_{i,\up}} + \avg{n_{i,\down}} , \\
    \avg{b_{ij,\s}} = \avg{\cd_{i,\s}\cn_{j,\s}} ,
\end{align}
\end{subequations}
and the $\avg{ij}$ summation denotes a sum over nearest-neighbors in which both the $(i,j)$ and $(j,i)$ bonds are counted.

In this work, we investigate spin and/or charge ordered phases at half band filling. Such phases are described by mean fields of the form
\begin{subequations}
\begin{equation}
    \avg{n_{i,\s}} = \frac{1}{2} + \delta n_\s \exp(\ii \bQ\cdot\bR_i) ,
\end{equation}
\begin{equation}
    \delta n_{\s} = \frac{1}{2} \left( \delta n + \underline{\tau}^3_{\s\s} m_z \right) ,
\end{equation}
\begin{equation}
    \avg{b_{ij,\s}} = b_{\s} + \delta b_\s \exp(\ii \bQ\cdot\bR_i),
\end{equation}
\end{subequations}
with $\bQ=(\pi,\pi)$ an ordering wave-vector and $\underline{\tau}^3$ the third Pauli matrix. 
We introduced the charge polarization $\delta n = (n_A - n_B)/2$, 
the staggered magnetization $m_z = (n_{A,\up} - n_{A,\down} - n_{B,\up} + n_{B,\down})/2$, 
and the (spatially homogeneous) spin projected bond charge $b_{\s}$ as well as its spatially modulated 
part $\delta b_\s$.
With $\H_{ext}$ now being expressed as a free fermion Hamiltonian, its diagonalization in Fourier space 
is straightforward and we can derive the free energy per lattice-site
\begin{align} \label{model:f_hf}
    \frac{F}{N_L} = & -\frac{1}{\beta N_L} \sump{\bk,\s,\nu} \ln\left[1+\exp(-\beta E^{HF}_{\bk,\s,\nu})\right] + \mu - \frac{U}{4} - 2V \notag\\
    & + \left(2V-\frac{U}{4}\right) \delta n^2 + \frac{U}{4} m_z^2 - 2V \sum_{\s} \left( b_{\s}^2 + \delta b_{\s}^2 \right) ,
\end{align}
where the primed sum is performed over the reduced BZ of the bipartite square lattice and 
$\beta=1/T$ is the inverse temperature. 
If we also restrict the hopping processes to nearest neighbors ($t_{ij}=-t$ if $i$ and $j$ are 
neighboring sites and $t_{ij}=0$ else), the eigenvalues of the one-body part of the Hamiltonian 
read
\begin{equation}
    E^{HF}_{\bk,\s,\nu} =  \frac{U}{2} + 4V -\mu + \nu \sqrt{ (t^{eff}_{\bk,\s})^2 + (\Delta^{HF}_{\s})^2 } ,
\end{equation}\label{eq:disp_hf}
with $\nu=\pm 1$. The effective dispersion read
\begin{equation}\label{eq:eff_disp_hf}
    t^{eff}_{\bk,\s} = 2(Vb_{\s}-t) ( \cos k_x + \cos k_y ) , \\
\end{equation}
and the band gaps
\begin{equation}
    2\Delta^{HF}_{\s} =  \left(U-8V\right)\delta n - \underline{\tau}^3_{\s\s}U m_z - 2\m .
\end{equation}
We can thus solve the self-consistent field equations for the introduced order-parameters,
$\delta n$, $m_z$ and $b_{\s}$, in the different phases (see \app{app:hf} for further details).

\subsection{The KRSB representation}

\subsubsection{Slave boson methods}

Building on Barnes' pioneering work \cite{barnes1976,barnes1977}, slave boson
representations for the most ubiquitous correlated electron models have been 
set up. In the specific case of the Hubbard model, the Kotliar-Ruckenstein (KR) 
representation \cite{kotliar1986} --- as well as its spin-rotation invariant and 
spin-and-charge-rotation invariant generalizations \cite{li1989,fresard1992} --- 
has been found to be of particular convenience and has been applied to a variety 
of cases.

Regarding its reliability, the KR representation has been shown to compare favorably 
with Quantum Monte Carlo (QMC) simulations. Specifically, for $U=4~t$ it could be shown that 
the slave-boson ground-state energy is larger than its QMC counterpart by less than $3\%$ 
\cite{lilly1990,fresard1991}. Additionally, very good agreement with QMC simulations on the location 
of the metal-to-insulator transition for the honeycomb lattice has been demonstrated 
\cite{fresard1995}. It should also be emphasized that quantitative agreement of the spin and 
charge structure factors of QMC and Density Matrix Embedding Theory were established 
\cite{zimmermann1997,dao2017,riegler2020}. 
Comparison with exact diagonalization has been performed, too. For values of $U$ larger than $4~t$, 
it has been obtained that the slave-boson ground state energy exceeds the exact diagonalization 
data by less than $4\%$ ($7\%$) for $U=8~t$ ($20~t$) and doping larger than $15\%$. 
This discrepancy decreases when the doping is lowered \cite{fresard1992b}. 
For small values of $U$ and close to half-filling, on the other hand, it has been shown that 
the KR representation and the HF approximation yield quantitatively similar energies 
\cite{fresard1991,igoshev2015,igoshev2016,steffen2017,igoshev2021}, but sizable differences arise 
for $U$ reaching half the band width.

Additionally, the slave boson approach exhibits several intriguing formal properties. 
First, the approach has been found to yield exact results in the large degeneracy 
limit \cite{fresard1992,florens2002}. Moreover, in the KR representation, the 
paramagnetic saddle-point approximation turns out to reproduce the Gutzwiller 
approximation~\cite{kotliar1986,gutzwiller1964}. It therefore inherits its formal 
property of obeying a variational principle in the limit of large spatial dimensions, 
where the Gutzwiller approximation and the Gutzwiller wavefunction \cite{gutzwiller1965} 
become identical \cite{metzner1988,metzner1989}. These properties are compelling 
evidence that the approach captures characteristic features of strongly correlated 
electron systems such as the suppression of the quasiparticle weight and the 
Mott-Hubbard/Brinkman-Rice transition to an insulating state at half filling under 
increasing on-site Coulomb interaction strength \cite{brinkman1970}. 

In addition to 
the aforementioned properties, slave boson representations possess their own gauge 
symmetry group, which allows one to gauge away the phase of one (or several depending 
on the specific representation) slave boson field. In the case of the KR 
representation, only the field associated to double occupancy remains complex, 
while the Lagrange multipliers are promoted to time-dependent fields 
\cite{fresard1992} as detailed below. Such a representation gives rise to real-valued 
boson fields that are thus free from Bose condensation. Their expectation values are
generically finite and can be well approximated in the thermodynamic limit via the 
saddle-point approximation. Corrections to the latter may be 
obtained when evaluating the Gaussian fluctuations \cite{lhoutellier2015,dao2017} 
and the correspondence between this more precise evaluation and the time-dependent 
Gutzwiller approach could recently be achieved --- though by means of an extension 
of the latter compared to its original formulation \cite{bunemann2013,noatschk2020}. 
While exact results may be obtained for, \textit{e.~g.}, a simplified single-impurity 
Anderson model, the Ising chain, or small correlated clusters \cite{fresard2001,
fresard2007,fresard2012b,dao2020}, this comes at the cost of more involved 
calculations. Here instead, we make use of the KR representation in the saddle-point 
approximation in order to address the different orders in the phase diagram of the 
considered extensions of the Hubbard model at half band filling, in the thermodynamic 
limit.

\subsubsection{The Kotliar and Ruckenstein representation}

In the KR representation, a doublet of pseudofermions $\{\fn_\up,\fn_\down\}$ 
and a set of four slave bosons $\{\en,\pn_\up,\pn_\down,\dn\}$ are introduced at each lattice site in order 
to reconstruct the Hilbert space of the model. The latter are tied to the four distinct atomic 
configurations: empty, singly occupied (with spin projection $\up$ or $\down$) and doubly 
occupied, respectively. As such, the introduced auxiliary operators generate redundant 
degrees of freedom which are to be discarded. This can be achieved by enforcing the following 
three constraints at each lattice site,
\begin{subequations}
\begin{align}
    0 &= \ed_i \en_i + \sum_\s \pd_{i,\s} \pn_{i,\s} + \dd_i \dn_i - 1 , 
    \label{model:constraint_a} \\
    0 &= \fd_{i,\s} \fn_{i,\s} - \pd_{i,\s} \pn_{i,\s} - \dd_i \dn_i 
    \quad\quad (\sigma=\uparrow,\downarrow). 
    \label{model:constraint_b}
\end{align}
\end{subequations}
The first constraint enforces completeness of the representation, 
while the second and third ones ensure a one-to-one correspondence between bosonic 
and pseudofermionic densities at each lattice site. 
This yields a representation in which the boson (pseudofermion) operators satisfy the 
canonical (anti) commutation relations. Under this constrained representation, 
the electron-density operators can be mapped onto bosonic or pseudofermionic operators,
\begin{equation}
    n_{i,\s} \rightarrow \fd_{i,\s} \fn_{i,\s} = \pd_{i,\s} \pn_{i,\s} + \dd_i \dn_i , 
\end{equation}
while the transition operators are mapped onto a combination of both bosonic and 
pseudofermionic operators
\begin{equation}
    \cd_{i,\s} \cn_{j,\s} \rightarrow \fd_{i,\s} \zd_{i,\s}  \zn_{j,\s} \fn_{j,\s} ,
\end{equation}
with
\begin{equation}
    \zn_{i,\s} = \ed_i L^{\phantom{X}}_{i,\s} R^{\phantom{X}}_{i,\s} \pn_{i,\s} 
    + \pd_{i,-\s} L^{\phantom{X}}_{i,\s} R^{\phantom{X}}_{i,\s} \dn_i .
\end{equation}
Multiple distinct choices of $L^{\phantom{X}}_{i,\s}$ and $R^{\phantom{X}}_{i,\s}$ 
yield equivalent representations for the Hubbard model when the constraints are exactly enforced. 
In practice, these factors are always chosen as
\begin{subequations}
\begin{align}
    &L^{\phantom{X}}_{i,\s} = \left( 1 -\pd_{i,\s} \pn_{i,\s} - \dd_i\dn_i \right)^{-1/2} , \\
    &R^{\phantom{X}}_{i,\s} = \left( 1 - \pd_{i,-\s} \pn_{i,-\s} - \ed_i \en_i \right)^{-1/2} .
\end{align}
\end{subequations}
such that the saddle-point approximation yields correct results ($\avg{z_{i,\s}}=1$) in the non-interacting limit \cite{kotliar1986}, the empty band limit $n\rightarrow 0$ and the filled band limit $n\rightarrow 2$. Furthermore, when the constraints \eq{model:constraint_a} and \eq{model:constraint_b} are exactly enforced on each site, the auxiliary bosonic operators act as projectors onto the physical (empty, singly, or doubly occupied) states of the lattice site. It is then straightforward to see that the on-site Coulomb interaction term in the Hamiltonian can be rewritten as
\begin{equation}
    \H_U = U \sum_i \dd_i \dn_i ,
\end{equation}
which is now a mere quadratic bosonic term. In the following, the intersite Coulomb 
term $\H_V$ is represented as purely bosonic, such that
\begin{equation}
    \H_V = \frac{V}{2} \sum_{\substack{\avg{\rm ij}\\ \s,\s'}} \left( \pd_{i,\s} \pn_{i,\s} + \dd_i \dn_i \right) \left( \pd_{j,\s'} \pn_{j,\s'} + \dd_j \dn_j \right) .
\end{equation}
The $\H_{\m}$ term is represented as a bosonic contribution as well,
\begin{equation}
    \H_{\m} = \sum_{i,\s} \m_i \left( \pd_{i,\s} \pn_{i,\s} + \dd_i \dn_i \right) .
\end{equation}

\subsubsection{Gauge fixing}

In the functional integral formalism, the partition function $Z$ of the system reads
\begin{equation}
    Z = \int_{-\pi/\beta}^{\pi/\beta} 
    \prod_i
    \frac{\d\lambda^{\vphantom{X}}_{i\vphantom{\up}}}{2\pi} 
    \frac{\d\lambda'_{i,\up}}{2\pi} \frac{\d\lambda'_{i,\down}}{2\pi} 
    \int \D[\phi^*,\phi] \e^{-\Omega},
    \label{model:path_integral}
\end{equation}
with 
$\phi_i (\tau) \equiv 
[ f_{i,\up}(\tau),f_{i,\down}(\tau),e_i(\tau),d_i(\tau),\pn_{i,\up}(\tau),\pn_{i,\down}(\tau) ]$
, where $f_{i,\s}(\tau)$ and $e_i(\tau),d_i(\tau),...$ now refer to time-dependent Grassmann 
and complex fields, respectively, 
while $\lambda_i,\lambda'_{i,\s} \in \mathbb{R}$ are the Lagrange multipliers enforcing the 
constraints \eq{model:constraint_a} and \eq{model:constraint_b}. 
In order to simplify the notation, the time dependence of the fields are not explicitly written in 
the remaining of this paragraph.

Following \rrefs{fresard1992,dao2020}, we gauge away the phases of three bosonic fields and express them as 
real amplitudes, while the remaining field --- most of the time chosen to be the $\dn$ field --- remains 
complex-valued. Simultaneously, the Lagrange multipliers $\lambda_i$ and $\lambda'_{i,\s}$ must be 
promoted to real time-dependent fields $\an_i$ and $\bn_{i,\s}$
\begin{subequations}
\begin{align}
    &\an_i \equiv \lambda_i + \partial_\tau \theta_i , \\
    &\bn_{i,\s} \equiv \lambda'_{i,\s} - \partial_\tau \chi_{i,\s} ,
\end{align}
\end{subequations}
where $\theta_i$ and $\chi_{i,\s}$ are the (time dependent) phase factors of the gauge 
transformation.

Within this gauge, the action $S = S_f + S_b$ of the path integral reads
\begin{equation} \label{eq:sf}
    S_f = \! \int_0^\beta \! \d\tau \! \sum_{i,\s} \! \left[ 
    f^*_{i,\s} \left( \partial_\tau - \mu + \ii \bn_{i,\s} \right) \fn_{i,\s} 
    + \sum_{j} t^{\phantom{0}}_{ij} f^*_{i,\s} z^*_{i,\s}  \zn_{j,\s} \fn_{j,\s} \! \right] ,
\end{equation}
for the mixed bosonic-pseudofermionic sector, and
\begin{equation}
\begin{split}
    S_b =& \int_0^\beta \d\tau \Bigg[ 
    \sum_i \ii \an_i \left( e^2_i + \sum_\s p^2_{i,\s} + \lvert \dn_i \rvert^2 - 1 \right) 
    + d^*_i \partial_\tau d^{\vphantom{*}}_i \\
    &- \sum_{i,\s} \ii \bn_{i,\s} n_{i,\s} 
    + U \sum_i \lvert \dn_i \rvert^2 + \frac{V}{2} \sum_{\avg{\rm ij}} n_i n_j + \sum_i \m_i n_{i} \Bigg] ,
\end{split}
\end{equation}
for the purely bosonic sector (all densities are understood to be expressed in 
terms of bosonic fields here and throughout).

\subsection{The saddle-point approximation}

The present study is performed at the saddle-point level of approximation, 
implying that the slave boson fields 
$\psi_i = [ \en_i,\Re{d_i},\Im{d_i},\pn_{i,\up},\pn_{i,\down},\bn_{i,\up},\bn_{i,\down},\an_i ]$ are averaged out such that they can be replaced by their time-independent 
expectation value $\left.\avg{\psi_i}\right|_{i\in s} \equiv \psi_s$ ($s=A,B$). 
One can thus explicitly integrate out the Grassmann fields and make use of the 
free fermion character of the KR representation (see \app{app:saddle_point}) to 
rewrite the fermionic sector of the grand potential
\begin{equation} \label{eq:sf0}
    \frac{\Omega_f}{N_L} = -\frac{1}{\beta N_L} \sump{\bk,\s,\nu} \ln 
    \left[ 1 + \exp (-\beta E_{\bk,\s,\nu}) \right] .
\end{equation}
If, again, we restrict the hopping to nearest-neighbors only, the quasiparticles dispersion is given by
\begin{equation}\label{eq:disp_sb}
    E_{\bk,\s,\nu} = \bmoy_\s - \mu 
    + \nu \sqrt{ ( \tilde{z}^2_{\s} t^{\vphantom{0}}_{\bk} )^2 + (\Delta^{SB}_{\s})^2} ,
    \quad \nu=\pm1 ,
\end{equation}
where we introduced,
\begin{subequations}
\begin{align}
&\bmoy_\s \equiv  \frac{1}{2} \left( \bn_{A,\s} + \bn_{B,\s} \right) , \\
&\Delta^{SB}_{\s} \equiv \frac{1}{2} \left( \bn_{A,\s} - \bn_{B,\s} \right) ,
\end{align}
\end{subequations} 
and the usual tight-binding dispersion of the square lattice
\begin{equation}
    t^{\vphantom{0}}_\bk = -2t \left( \cos k^{\vphantom{y}}_x + \cos k^{\vphantom{x}}_y \right) .
\end{equation}
The renormalization factors now read
\begin{subequations}
\begin{equation}
    \tilde{z}_{\s} = \sqrt{\zn_{A,\s} \zn_{B,\s}} ,
\end{equation}
with
\begin{equation}
    \zn_{s,\s} = \frac{\en_s \pn_{s,\s} + \pn_{s,-\s} \dn_s}{\sqrt{n_{s,\s}\left(1-n_{s,\s}\right)}}.
\end{equation}
\end{subequations}
The purely bosonic contribution to the grand potential is given by
\begin{equation} \label{eq:sb0}
\begin{split}
    \frac{\Omega_b}{N_L} &= \frac{\an_A}{2} \left( e^2_A + \sum_\s p^2_{A,\s} + \left|d_A\right|^2 - 1 \right) \\
    &+ \frac{\an_B}{2} \left( e^2_B + \sum_\s p^2_{B,\s} + \left|d_B\right|^2 - 1 \right) 
    - \sum_\s \bmoy_\s \avg{n_{\s}} \\
    &- \sum_\s \frac{\Delta^{SB}_{\s}+\m}{2} ( \delta n + \underline{\tau}^3_{\s\s} m_z ) 
    + \frac{U}{2} \left( \left|d_A\right|^2 + \left|d_B\right|^2 \right) \\
    &+ 2V ( \avg{n}^2 - \delta n^2 ) ,
\end{split}
\end{equation}
where the averaged values and order parameters are defined identically to their 
counterparts in the HF approximation scheme.
Let us note that we can already notice that the $\Delta^{SB}_{\s}$ and $\m$ terms 
control the relative filling of the sublattices, as they couple to the difference 
in density between these two sublattices.
For more details, see \app{app:saddle_point} in which the saddle point equations (SPE) are derived.

\begin{figure*}
    \centering
    \includegraphics[width=.92\textwidth]{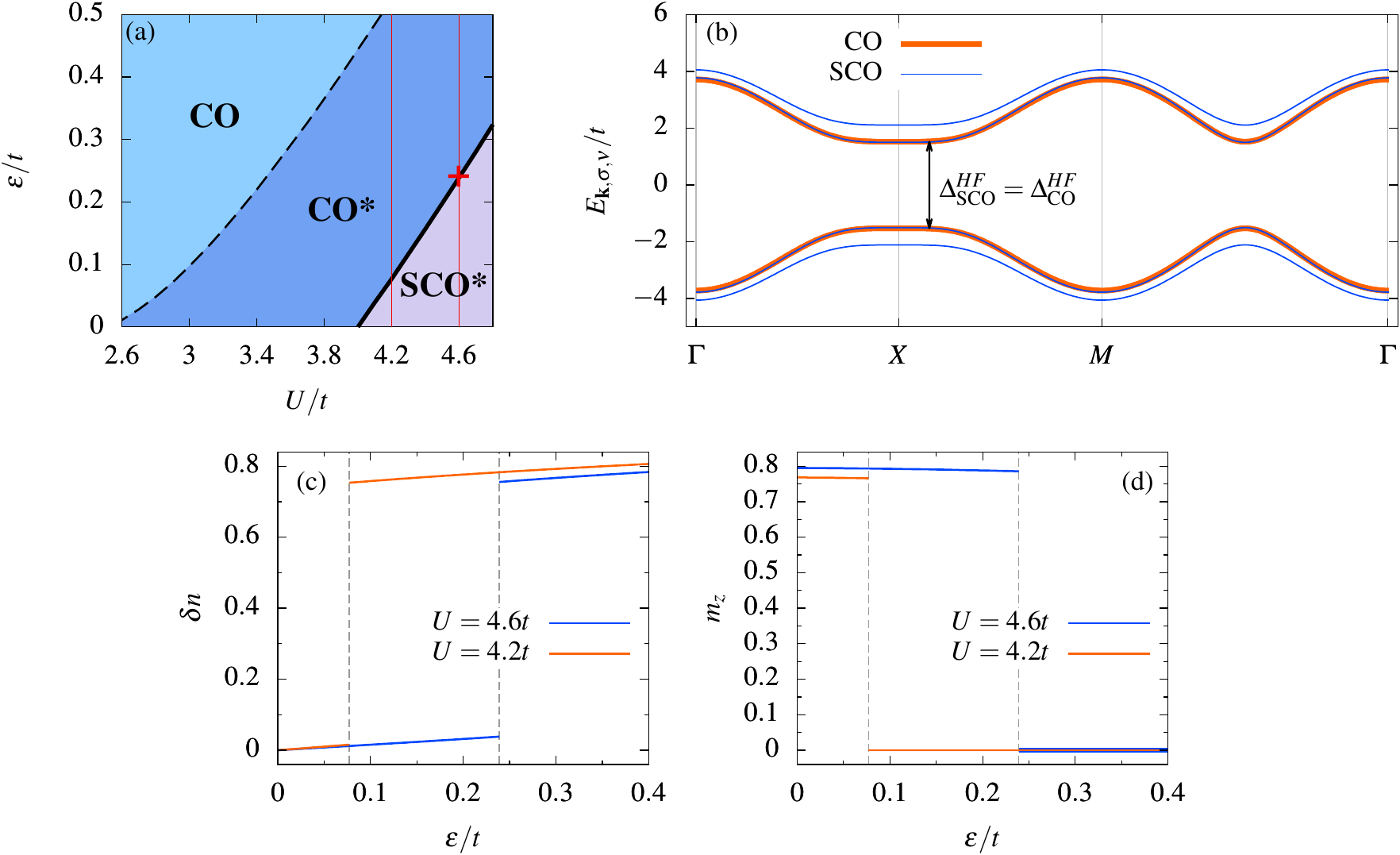}
    \caption{
    (a) Zero temperature phase diagram of the half-filled extended Hubbard model in the $(U, \m)$ plane
    for $V=t$ within the HF approximation.
    It comprises a CO region and a SCO region.
    The phase boundary between the two is denoted by the solid black line.
    The dashed line corresponds to the SCO end line, along which the SCO solution vanishes.
    In the CO${}^*$ (SCO${}^*$) region the CO and SCO phases coexist with the CO (SCO) one lower in energy.
    The red cross (lines) indicates the parameters (ranges) used for subsequent plots in this figure. 
    (b) Band structure for the pseudofermions at a SCO-CO transition point. 
    The arrows are indicative of the gaps, which are measured at the $X$ point. 
    Parameters: $U=4.2~t$, $V=t$, and $\m=0.24~t$.
    (c) $\delta n$, (d) $m_z$ as functions of $\m$. Parameters: $T=0$, $V=t$, $U=4.2~t$ and 
    $4.6~t$. The gray dashed lines denote transition points.
    }
    \label{fig:HF_1}
\end{figure*}

\section{Results with the HF approximation} \label{sec:res_hf}

In this section, we investigate the half-filled, zero temperature phase diagram of the above introduced 
Hubbard model extended by an intersite Coulomb repulsion and a SLP. 
Results for the half-filled Hubbard model within a 
similar inhomogeneous crystal field ($\H=\H_{Hub}+\H_{\m}$) have been obtained in 
one and two dimensions using Lattice Density Functional Theory 
\cite{saubanere2009,saubanere2011}. On the square lattice, these computations 
highlighted a continuous phase-transition from the charge-density wave (CDW) 
phase to a spin-ordered state as the ratio $U/\m$ is increased. On the other 
hand, the $t$--$U$--$V$ model ($\H=\H_{Hub}+\H_{V}$) has been thoroughly 
investigated in the literature. Among the features of this model which 
attracted great interest is the CDW to spin-density wave (SDW) phase 
transition at half-filling. The latter has been studied analytically by means 
of perturbation theory \cite{vanDongen1994a,vanDongen1994b} or in the slave 
boson saddle-point approximation \cite{wolff1983,deeg1993} and numerically 
from Quantum Monte Carlo (QMC) calculations in one \cite{hirsch1984,segpunta2002} 
or two \cite{zhang1989} dimensions. More recently, the Dynamical Cluster 
Approximation (DCA) has been applied to the square lattice 
\cite{terletska2017,paki2019}, yielding a detailed phase diagram of the model 
at finite temperature in the $U$--$V$ parameter space. To the best of our 
knowledge, however, no study that simultaneously takes both extensions 
of the Hubbard model into account has been performed.

The proposed origin of the SLP does not justify the investigation of values of $\m$ larger 
than the energy scale $t$.
Yet, we show below that small values of $\m$ may change the nature of the ground state.

\begin{figure*}
    \centering
    \includegraphics[width=.92\textwidth]{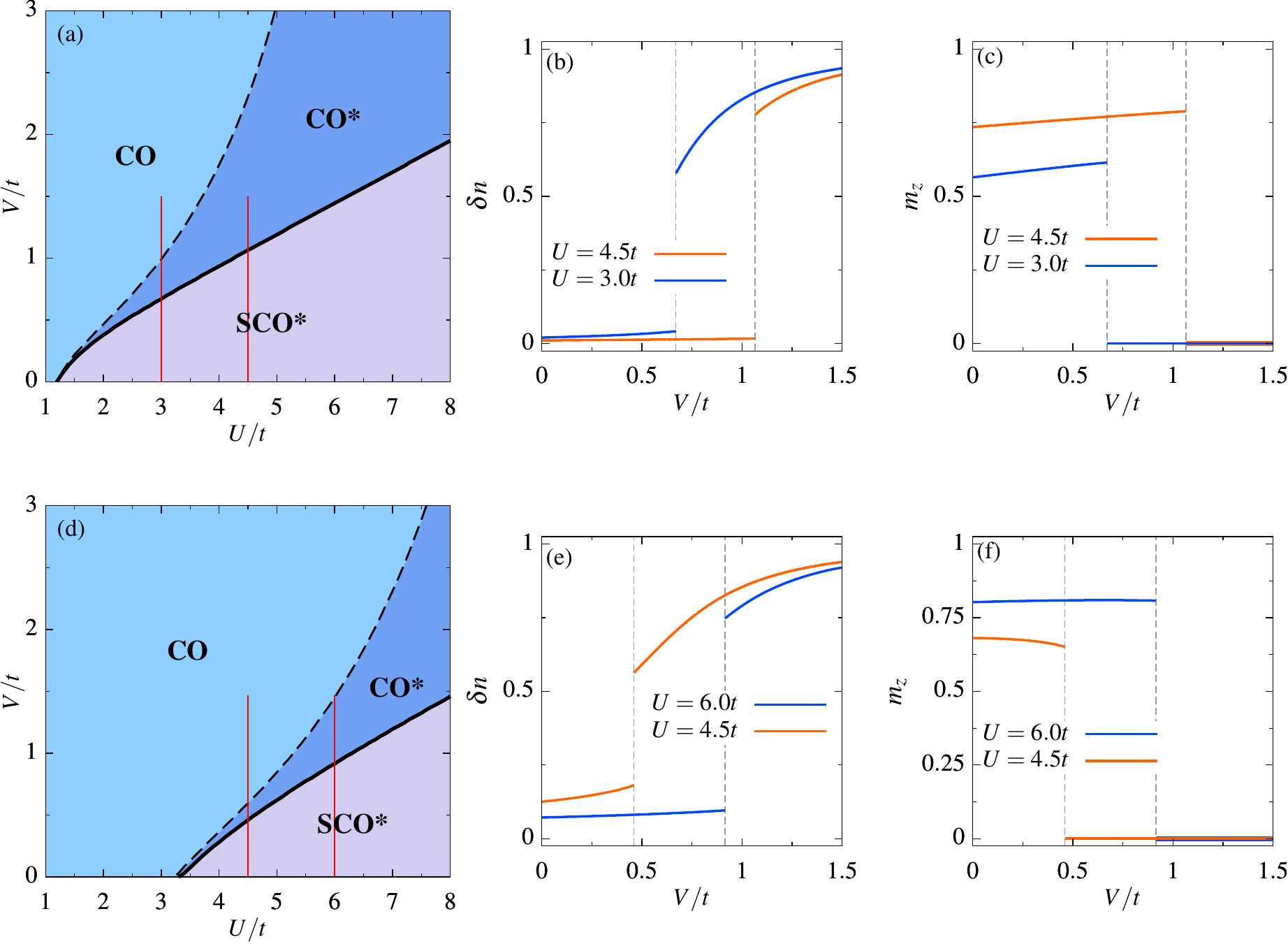}
    \caption{
    Zero temperature phase diagram of the half-filled extended Hubbard model in the $(U, V)$ plane
    for (a) $\m=0.1~t$, and (d) $\m=t$ within the HF approximation.
    It comprises both a CO and a SCO region.
    The phase boundary between the two is denoted by the solid black line.
    In the CO${}^*$ (SCO${}^*$) region the CO and SCO phases coexist with the CO (SCO) one lower in energy.
    The end-lines of the two phases are denoted by dashed lines.
    The red lines indicate the parameters ranges used for subsequent plots in this figure.
    (b) $\delta n$ and (c) $m_z$ as functions of $V$. Parameters: $T=0$, $\m=0.1~t$, $U=3~t$ and $4.5~t$. 
    (e) $\delta n$ and (f) $m_z$ as functions of $V$. Parameters: $T=0$, $\m=t$, $U=4.5~t$ and $6~t$.
    }
    \label{fig:HF_2}
\end{figure*}

\subsection{Influence of the local potential}

The zero temperature phase diagram of the three parameter half-filled extended model may be 
computed through minimization of \eq{model:f_hf} with respect to $\delta n$, $m_z$ and $b_\s$.
Yet, it may be superfluous to scan this three dimensional parameter space and we stick to a series of
two dimensional subspaces by attributing a fixed value to the third parameter, aimed at unraveling 
the most prominent features of the model. We start our investigations maintaining $V$ fixed, and we 
compute the phase diagram, together with the variations of the order parameters in order to 
establish the nature of the phase transitions.

The resulting phase diagram is displayed in \fig{fig:HF_1}(a) for fixed $V=t$. 
It comprises two phases, namely a charge ordered (CO) phase and a spin-and-charge ordered (SCO) phase.
The CO* phase corresponds to a region of the phase diagram in which the ground state is CO but 
SCO-type solutions to the HF self-consistent field equations also exist and usually lie close in 
energy. The SCO* phase is defined similarly, with SCO being the stable phase. The CO and CO* regions 
are thus separated by the SCO end-line, which corresponds to the line in parameter space along 
which SCO solutions vanish and only CO solutions remain.
The phase diagram is in line with expectation: while increasing $U$ tends to stabilize the SCO phase, 
increasing $\m$ promotes the CO one.
For comparatively large $U$ ($U\gtrsim 4~t$), the SCO phase extends down to $\m=0$, in which case a pure 
SDW phase is stabilized. This SDW phase entails no charge ordering despite of the finiteness of $V$.

\fig{fig:HF_1}(b) shows the band structure of the two phases at the transition point $U=4.6~t$, 
$V=t$ and $\m=0.24~t$. In the CO phase, the band structure is simple, since 
$\Delta^{HF}_{CO}=(U/2-4V)\delta n - \m$ for both spin branches. 
This results in the usual two-band system reminiscent of the CDW phase of the $t$--$U$--$V$ model. 
In the SCO phase, however, we have $\Delta^{HF}_{SCO,\up}=(U/2-4V) \,\delta n - m_z U/2 - \m$ and 
$\Delta^{HF}_{SCO,\down}=\Delta^{HF}_{SCO,\up} + m_z U$. Hence, the spectrum entails four branches, 
in contrast to the usual two branches of the antiferromagnetic (AFM) phase of the $t$--$U$--$V$ model.
Since we work on a half-filled lattice, the four branches are centered around zero and, at zero 
temperature, the $\nu=-1$ bands are entirely filled while the $\nu=+1$ ones are empty. 
We also note that the difference in the gaps of the two spin branches, added to the fact that these branches 
are centered around the same value, yields a symmetry broken phase in which one of the spin branches is 
higher in energy than the other one. The choice of which branch is higher or lower in energy remains, however, 
purely arbitrary and is fixed by the sign of $\delta n$ and $m_z$ in the SCO solutions. In reality, all four 
different $(\pm \delta n,  \pm m_z)$ solutions are degenerate and can only be distinguished by their 
internal parameters.
Moreover, since the leading contributions to the Fermi integrals in this regime come from the 
band minima, we expect the energy of both phases to become nearly degenerate when the minima 
of the CO $\nu=-1$ band coincide with the highest SCO $\nu=-1$ band. 
This is precisely what happens at the shown transition point, yielding a SCO-CO phase transition. 
We can also expect this transition to be discontinuous, as we observe a difference between 
$\Delta^{HF}_{SCO,\up}$ and $\Delta^{HF}_{SCO,\down}$  of order $2~t/3$, implying that the 
magnetization in this phase is still large at the transition point and will have to drop down to 
zero in the CO phase.

In \fig{fig:HF_1}(c) and \fig{fig:HF_1}(d), the $\m$-dependence of the order parameters at $T=0$, 
$V=t$ and for fixed values of $U=4.2~t$ and $U=4.6~t$ are shown. 
We first notice that the charge ordering is more pronounced in the CO phase than it is in the SCO 
one. Consequently, both phases are still completely distinct at the phase transition, 
yielding a discontinuous jump of the order parameters characteristic of a first order phase 
transition. For smaller values of $U$, however, the antiferromagnetic ordering of the spins is 
somewhat weakened, resulting into slightly less marked discontinuities. 
Nevertheless, no continuous variation of the order parameters at the transition point has been found 
in the finite $V$ regime in the relevant $U$ and $\m$ ranges. 
The transition is therefore discontinuous.

\begin{figure*}
    \centering
    \includegraphics[width=.92\textwidth]{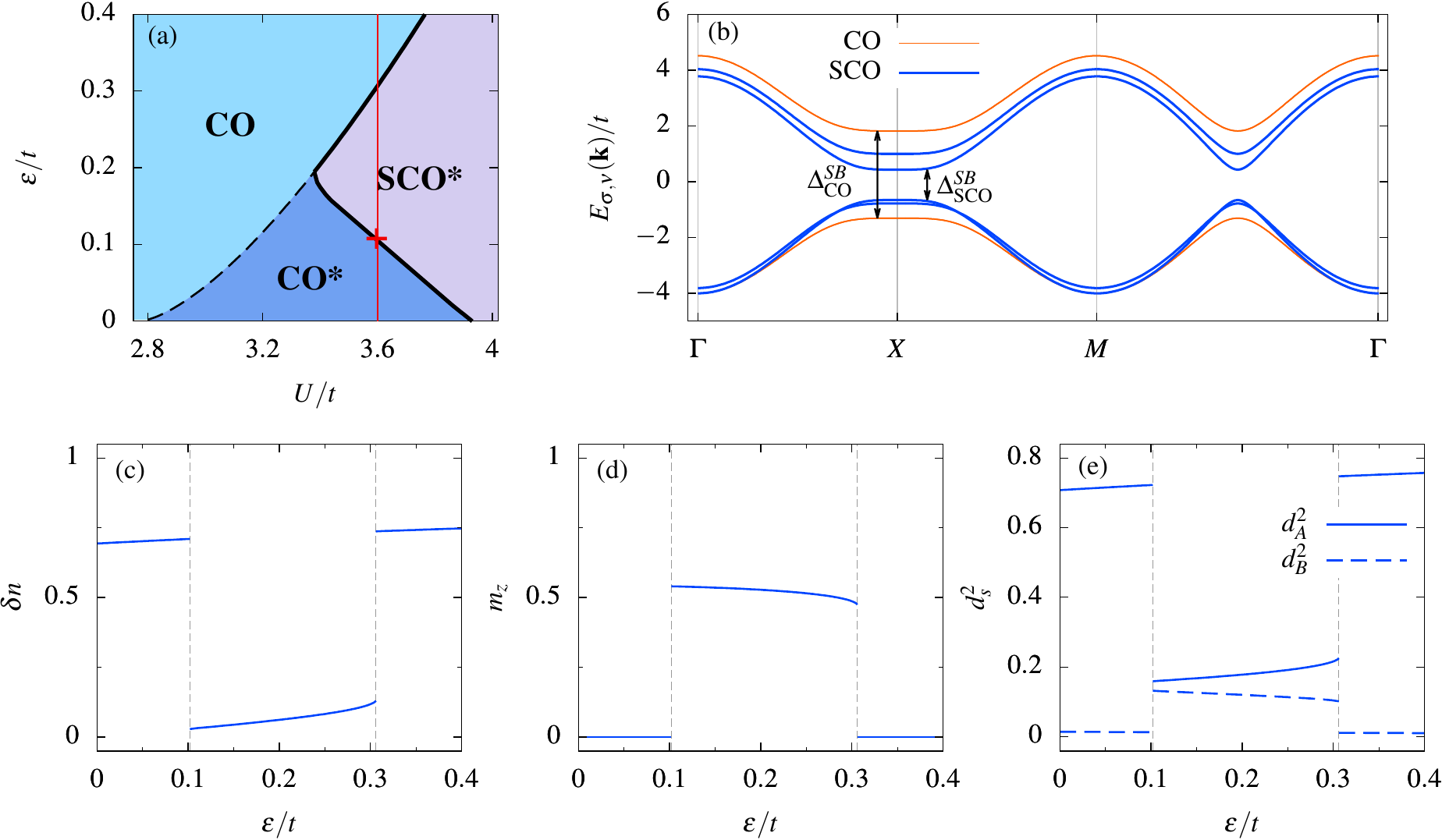}
    \caption{
    (a) Zero temperature phase diagram of the half-filled extended Hubbard model in the $(U, \m)$ 
    plane for $V=t$, within KRSB formalism. 
    It comprises a CO region and a SCO region. 
    The phase boundary between the two is denoted by the solid black line. 
    The dashed line
    corresponds to the SCO end line, along which the SCO solution to the SPE
    vanishes. In the CO* (SCO*) region, both solutions exist but the CO (SCO)
    solution is lower in energy.
    The red cross (line) indicates the parameters (range) used for subsequent plots in this figure.
    (b) Band structure for the pseudofermions at a SCO-CO transition point. 
    The arrows are indicative of the gaps, which are measured at the $X$ point. 
    Parameters: $U=3.6~t$, $V=t$, and $\m=0.1~t$. 
    (c) $\delta n$, (d) $m_z$, (e) $d^2_A$ and $d^2_B$ as functions of $\m$ for fixed values of 
    $T=0$, $V=t$ and $U=3.6~t$. The gray dashed lines denote transition points.}
    \label{fig:tUVW_1}
\end{figure*}

\subsection{Dependence on the screened Coulomb interaction}

We now switch to the $U$-$V$ phase diagram. 
It is presented for $T=0$ and $\m=0.1~t$  as well as $\m=t$ in \fig{fig:HF_2}(a) 
and \fig{fig:HF_2}(d), respectively. 
Both phases are still separated by a first order transition line but, here, 
it merges with the SCO end-line at weak couplings. At stronger couplings, the transition line at 
fixed $\m$ becomes linear in $U$: $V(U) \simeq U/4$ for $U,V \gg \m$. 
This $V(U) \propto U/4$ dependence is evocative of the already reported $V=U/4$ CDW-SDW transition 
line of the half-filled $t$--$U$--$V$ model 
(see \rrefs{vanDongen1994a,vanDongen1994b,wolff1983,deeg1993,hirsch1984,segpunta2002,zhang1989,terletska2017,paki2019}) 
that we address below. 
This is to be expected as, in a coupling regime in which $\m$ becomes negligible, one should 
recover results qualitatively similar to those of the $t$--$U$--$V$ model (in which $\m=0$). 
This also explains why the beginning of this linear regime is shifted towards stronger couplings 
as $\m$ is increased from $0.1~t$ [\fig{fig:HF_2}(a)] to $t$ [\fig{fig:HF_2}(d)].

In \fig{fig:HF_2}(b) and \fig{fig:HF_2}(c) the zero temperature variations of the order parameters 
are presented, for representative parameter sets, as functions of V. 
We set $\m=0.1~t$, together with $U=3~t$ and $U=4.5~t$. 
In \fig{fig:HF_2}(e) and \fig{fig:HF_2}(f) the same variations are presented but for fixed $\m=t$, 
and for values of $U=4.5~t$ and $U=6~t$. We obtain a single SCO-CO transition for each parameters set, 
for $V=0.67~t$, $1.14~t$, $0.46~t$ and $0.91~t$, respectively. 
The order parameters again vary discontinuously at these transitions. 
We also note that this discontinuity is preserved at weaker and stronger couplings. 
We see, however, that the discontinuity is softened for larger values of $\m$ and/or smaller values 
of $U$. This owes to the stronger antiferromagnetic ordering of the spins and weaker charge ordering 
at large values of $U/\m$, yielding a sharper drop (increase) in $m_z$ ($\delta n$) at the SCO to CO 
transition. Combining these results with the ones of the previous section, we conclude that, 
in the HF approximation, the SCO-CO phase transition occurring at zero temperature, and half-filling,
is discontinuous. 
In addition, the phase diagram exhibits a phase boundary between the CO* and SCO* in a large range of 
couplings, which may be suitable for resistive switching.

\section{Results with the KRSB representation}\label{sec:res_krsb}

As highlighted in, \textit{e.g.}, \rref{steffen2017}, the HF approximation and the KRSB representation of 
extended (or not) Hubbard models are in good quantitative agreement in the weak-coupling regime. 
However, one can hardly expect the HF approximation to yield quantitatively appreciable results in the 
intermediate-to-strong coupling regime due to the inherent inability of such mean field techniques to 
account for strong correlations. On the other hand, as outlined earlier, slave boson techniques have 
proved their efficiency in the stronger coupling regimes \cite{fresard2012,fresard1991,fresard2022,
riegler2020,igoshev2015,igoshev2016,igoshev2021}. The KRSB representation thus appears as a convenient tool 
to assess for the relevance of the correlations in the strong coupling regime of the model and we shall use 
it to that aim in this section.

\subsection{Influence of the local potential}

Let us now show the results following from the solution of the saddle-point equations \eq{eq:spe} 
and discuss them in the next two paragraphs.
Emphasized is the competition between the different orders in the phase diagram. An 
identification of the lowest energy solution --- with the free energy per 
lattice site $F=\Omega/N_L+\mu$ --- allows us to determine the stable phase for 
given values of the parameters and we study the variations of the order 
parameters $\delta n$, $m_z$ and $d^2_s= \avg{n_{i,\up}n_{i,\down}}|_{i\in s}$ in order to confirm 
the nature of the different phase transitions.

\begin{figure*}
    \centering
    \includegraphics[width=.92\textwidth]{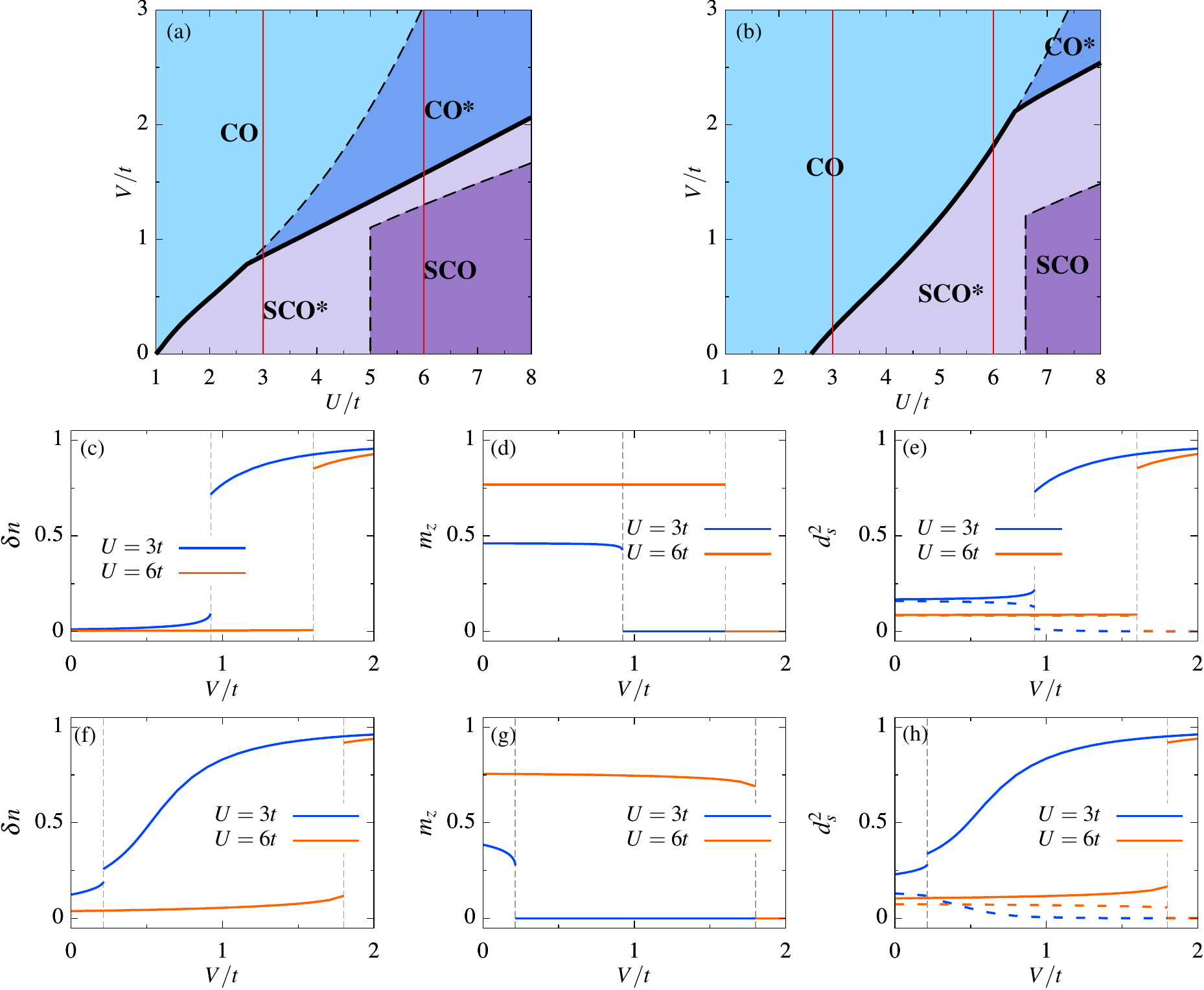}
    \caption{
    Zero temperature phase diagram of the half-filled extended Hubbard model in the $(U, V)$ plane
    for (a) $\m=0.1~t$, and (b) $\m=t$, within KRSB formalism. It comprises both a CO and a SCO
    region. The phase boundary between the two is denoted by the solid black
    line. In the CO${}^*$ (SCO${}^*$) region the CO and SCO phases coexist
    with the CO (SCO) one lower in energy. The end-lines of the two phases
    are denoted by dashed lines. 
    The red lines indicate the parameter ranges used for subsequent plots in this figure.
    (c) $\delta n$, (d) $m_z$, (e) $d^2_A$ and $d^2_B$ as
    functions of $V$. Parameters: $T=0$, $U=3~t$ and $\m=0.1~t$. 
    (f) $\delta n$, (g) $m_z$, (h) $d^2_A$ (solid line) and $d^2_B$ (dashed line) as
    functions of $V$. Parameters: $T=0$, $\m=t$, $U=3~t$ and $6~t$. 
    }
    \label{fig:tUVW_2}
\end{figure*}

The resulting phase diagram is displayed in \fig{fig:tUVW_1}(a) as a function of $U$ and $\m$ at zero 
temperature and fixed $V=t$. 
It comprises a CO and a SCO region, separated by a phase boundary across which a first order phase 
transition occurs. 
The lowermost ($\m\lesssim 0.2~t$) segment of this transition line corresponds to the 
$F_{CO}=F_{SCO}$ 
degeneracy line while the uppermost ($\m\gtrsim 0.2~t$) segment lies along the SCO end-line. 
This is in contrast to the HF phase diagram in $U$-$\m$ space, in which the phase boundary only 
consists of the CO-SCO degeneracy line. 
Above the SCO end-line, the CO phase is stabilized by default, irrespective of its energy. 
Below this line, for lower couplings ($U \lesssim 3.8~t$), we find a dome in the CO phase in which 
the SCO solution still exists and lies close in energy. 
The SCO phase is thus meta-stable under this dome, implying coexistence of both phases in this 
regime. 
In addition, a pure SDW phase is stabilized for $\m=0$ and $U \gtrsim 3.5~t$ as already encountered in 
the HF phase diagram \fig{fig:HF_1}(a).

\fig{fig:tUVW_1}(b) presents the band structure of the two phases at a given 
transition point $U=3.6~t$, $V=t$ and $\m =0.102~t$. 
In the CO phase, $\mueff_\up=\mueff_\down$ (with $\mueff_\s=\mu-\bmoy_\s$) and 
$\Delta^{SB}_{\up}=\Delta^{SB}_{\down}$, resulting into the two spin branches being equivalent. 
Besides, no such simplification holds in the SCO phase, leading to four distinct bands.
Analogously to what has been discussed in \Sec{sec:res_hf}, the SCO solution has a broken spin-rotation 
symmetry, with two inequivalent spin branches. 
As in the HF approximation, the lowest and highest energy branches are fixed 
by the sign of $\delta n$ and $m_z$, while the four different $(\pm \delta n, \pm m_z)$ solutions 
remain degenerate.
Since we are at half-filling, the Lagrange multipliers and 
the chemical potential satisfy $|\mueff_\s| < |\Delta^{SB}_{\s}|/2$, such that only the 
$\nu=-1$ bands are filled, while the $\nu=+1$ bands remain empty. Looking at 
these lower bands, we see that --- at this transition point in $U$-$V$-$\m$ 
space --- the minima of the bands in the SCO phase overlaps with its counterpart in the CO phase. 
Varying the parameters towards the CO (SCO) stability region of the phase 
diagram results in the lowering of the minima of the CO (SCO) lower band. 
Since we are at zero temperature, the contributions to the Fermi integrals 
largely follows from these lower bands minima. The stable phase for a given 
set of parameters can thus be suggested by inspection of the latter. Besides, 
$\Delta^{SB}_{CO}$ markedly differs from $\Delta^{SB}_{SCO}$ even though both phases 
are degenerate in energy. Hence, the value of the gap may not be used to 
predict the nature of the ground state.

The variation of the order parameters is given in \fig{fig:tUVW_1}(c), \fig{fig:tUVW_1}(d), 
and \fig{fig:tUVW_1}(e) as a function of $\m$ and for fixed values of $T=0$, $U=3.6~t$ and $V=t$. 
We notice that the local charge and pair modulations are substantially larger in the CO phase than 
in the SCO one. 
For these parameters, two consecutive transitions between the two phases occur as $\m$ is 
increased ($\m\approx 0.1~t$ and $\m\approx 0.3~t$). 
We see that, at the transition points, all order parameters vary discontinuously. 
This implies the CO-SCO transition to be first order. 
At finite $V$, no region of parameter space has been found to exhibit a continuous variation of 
the order parameters across the phase boundary. 
However, the discontinuity in the variation of the order parameters at the transition points 
become less important as $\m$ and $V$ increase or as $U$ decreases. 
As detailed in the next paragraph, the SCO phase is restricted to small values of $V/U$, 
while large values of $\m/U$ are irrelevant due to the proposed origins for the $\m$-term. 
This shows that no continuous CO-SCO transition is possible in the parameter range of interest.

\subsection{Dependence on the screened Coulomb interaction}

Turning now to the explicit dependence on $V$ of the phase diagram, we present in 
\fig{fig:tUVW_2}(a) and \fig{fig:tUVW_2}(b) the zero-temperature phase diagram of the model as a 
function of $U$ and $V$, for $\m=0.1~t$ and $\m=t$. 
At weak couplings, the CO-SCO phase boundary merges with the SCO end-line [uppermost segment of the 
phase boundary in \fig{fig:tUVW_2}(a)]. 
At stronger couplings, however, the transition line between the two phases digresses from the SCO 
end-line and a $F_{CO}=F_{SCO}$ degeneracy line begins [lowermost part of the phase boundary in 
\fig{fig:tUVW_2}(a)]. 
In this regime, the transition line gradually becomes linear, such that it can ultimately be 
parametrized by $V=U/4+c(\m)$ for $U,V \gg \m$, with $c(\m)$ independent of $U$ and $V$. 
We computed $c(\m=0.1~t)\approx 0.06~t$ and $c(\m=t)\approx 0.53~t$ for these two phase diagrams. 
This strong coupling behavior is in qualitative disagreement with the one observed in the HF 
approximation. Indeed, the staggered potential appears to favor the stabilization of the SCO phase 
at large $U$ while, in the HF approximation, the SCO-CO transition line lies slightly below the 
$V=U/4$ line at strong coupling --- evidencing the role of correlations in the mechanisms 
underlying this phase transition in the strong coupling regime.
Let us also notice the vertical part of the CO end-line at $U=5~t$ and $6.2~t$ for $\m=0.1~t$ and $t$, 
respectively.
This follows from the fact that, for $U$ larger than these values, the charge order is 
generated by $\m$ alone and values of $V$ smaller than $t$ and $1.2~t$, respectively, do not 
yield additional charge order in the CO phase. 
In the $\m=0$ case, as discussed below, this would result in a disordered paramagnetic phase, 
which is prohibited here due to the minimal charge ordering imposed by $\m$.

The zero temperature variations of the order parameters as a function of $V$, for fixed $U=3~t$ and $U=6~t$ 
are shown in \fig{fig:tUVW_2}(c), \fig{fig:tUVW_2}(d) and \fig{fig:tUVW_2}(e) for $\m=0.1~t$, and 
in \fig{fig:tUVW_2}(f), \fig{fig:tUVW_2}(g) and \fig{fig:tUVW_2}(h) for $\m=t$. 
Along these paths in parameter space, there is a single CO-SCO transition at $V\approx 0.9~t$ for 
$\m=0.1~t$ and at $V\approx 0.2~t$ for $\m=t$. 
The order parameters vary discontinuously at this transition. 
Such jumps of order parameters happen as well in the linear regime of the transition line. 
We see that, for smaller values of $\m$, the discontinuity is enhanced. 
We also observe such an enhancement when $U$ is increased. 
This is explained by the fact that the SCO configurations become increasingly 
antiferromagnetically ordered and decreasingly charge ordered, and conversely for smaller values 
of $U$ and larger values of $\m$. 
Additionally, we note that small values of $\m$ with respect to the energy scales of $U$ and $V$ 
($\m=0.1~t$ compared to $U \geq t$ and $V\sim t$) are sufficient to establish strongly 
discontinuous transitions. 
For reasons outlined above, values of $\m$ larger than the energy scale $t$ have not been 
investigated. 
Moreover, in the small $U$ regime, the SCO phase is not stabilized, even for arbitrarily small 
values of $V$ if $\m$ is of the order of $t$. 
As a consequence to this, no transition point has been found to correspond to a continuous 
transition at finite $\m$. 
In addition to the statement that no second order transition occurs in $U$--$\m$ space for finite 
$V$, this leads to the conclusion that the CO-SCO transition at zero temperature is systematically 
of first order.
Furthermore, phase coexistence, suitable for resistive switching, has been established in a large 
domain of the phase diagram.

The way the SCO solution ceases to exist is reminiscent of the $U_{c1}$ and $U_{c2}$ lines of 
the two-band model in its $U$-dependence as described in the suitably extended KR representation.
There, for densities sufficiently close the half-filling and finite $J_H$, the quasiparticle 
residue acquires a vertical tangent and so does the free energy \cite{fresard2002,fresard2022}.
In other words, these singular points mark the end point of the insulating, resp. metallic 
solutions.
In the current case, the staggered magnetization of the SCO phase acquires a vertical tangent 
in its $V$-dependence (see \fig{fig:tUVW_2}), too, which is reflected in the $V$-dependence of 
the free energy as well.
Hence, this singular point denotes the end point of the SCO phase which, depending on the 
parameter values, may correspond to the phase transition.
For instance, this is the case for $t\lesssim U \lesssim 2.8~t$ with $\m=0.1~t$, while for
$\m=t$, this happens for $2.5~t \lesssim U \lesssim 6.5~t$.
For larger $U$-values, the phase transition does not correspond to the end point of the 
SCO phase any longer.

\subsection{Resistive switching}

Let us here briefly consider temperature driven SCO-CO phase transitions. They take place at temperatures 
scaling with the gap between the empty and occupied quasi-particle bands, which itself scales with $U$ 
when it is the largest energy into play. In that case, the largeness of the gap implies that the band 
filling is mostly independent of $T$ at small temperatures, leading to a high transition temperature. 
Yet, the situation changes in the intermediate coupling regime. 
For instance, we performed calculations with $U=4.2~t$, $V=t$ and $\m=0.1~t$, \textit{i.~e.} 
in the SCO phase close to its boundary, and we obtained the temperature 
at which the magnetic order melts to be close to 0.35~t while the SCO-CO transition was obtained at 
a lower $T_{SCO-CO}\simeq0.2~t$. 
Recalling that the hopping amplitude in transition metal oxides is widely 
accepted to be in the $0.1$ eV to $0.3$ eV range, it renders $T_{SCO-CO}$ experimentally 
accessible.

\begin{figure}
    \centering
    \includegraphics[width=.48\textwidth]{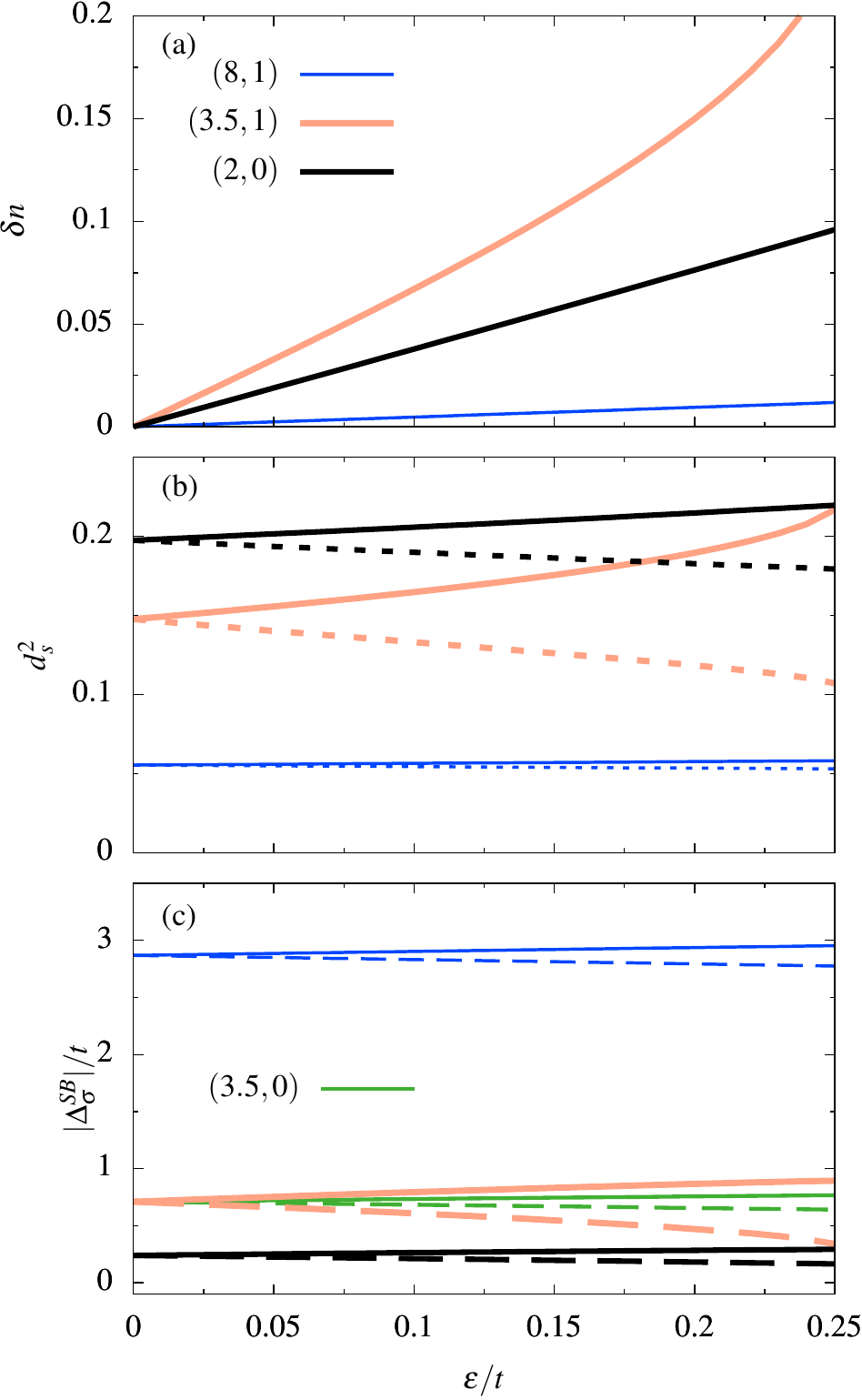}
    \caption{
    (a) $\delta n$, (b) $d^2_A$ (solid line) and $d^2_B$ (dashed line), (c) $|\Delta^{SB}_{\up}|$ 
    (solid line) and $|\Delta^{SB}_{\down}|$ (dashed line) in the SCO phase as functions of $\m$ for 
    different values of the couple $(U,V)$. These order parameters have been computed within KRSB 
    formalism. Parameters: $T=0$, $(U/t,V/t)=(8,1)$ (blue), $(3.5,1)$ (orange), $(2,0)$ (black) 
    and $(3.5,0)$ (green) in the last panel.
    }
    \label{fig:tUVW_limit}
\end{figure}

\begin{figure*}
    \centering
    \includegraphics[width=.92\textwidth]{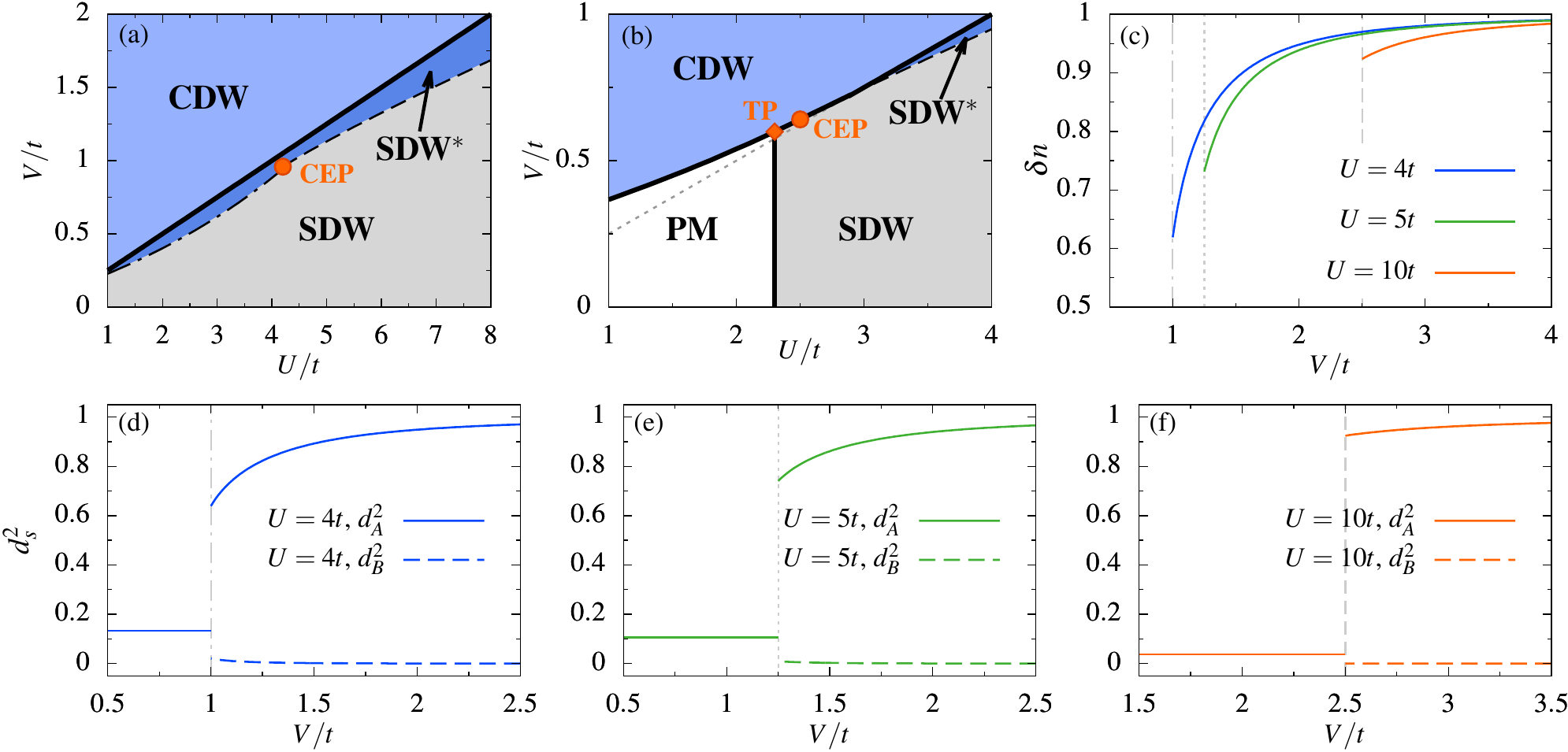}
    \caption{
    KRSB phase diagram of the half-filled $t$--$U$--$V$  model at (a) zero
    temperature and (b) $T=t/6$ in the $(U, V)$ plane. The phase boundaries
    are denoted by solid black lines. The dashed lines correspond to the CDW
    end-line, while the dash-dotted line denotes the PM-CDW instability
    line. The PM-CDW-SDW triple point (TP) is denoted by an orange diamond. The PM-CDW instability critical end-point (CEP) is denoted by an orange circle. 
    In the SCO${}^*$ region the CDW and SCO phases coexist,
    with the latter one being lower in energy. 
    (c) $\delta n$ as a function of $V$ at $T=0$ for $U=4~t$, $U=5~t$ and $U=10~t$. (d), (e) and (f) show $d^2_A$ and $d^2_B$ as a function of $V$ at $T=0$ and for $U=4~t$, $U=5~t$ and $U=10~t$, respectively. The gray lines correspond to the SDW-CDW transition points for the different values of $U$.}
    \label{fig:tUV}
\end{figure*}

\section{The \texorpdfstring{\MakeLowercase{t}--U--V}{t--U--V} limit}\label{sec:tuv_krsb}

\subsection{Zero temperature results}

As outlined in \Sec{sec:res_krsb}, the SCO phase is not observed in the $\m=0$ limit. 
The reason is that the SCO solution continuously becomes a pure SDW phase as 
$\m\to 0$. As can be seen from \fig{fig:tUVW_limit}(a), in this limit, we find that 
in the SCO solutions, $\delta n$ goes to zero for all values of $U$ and $V$. A 
similar effect is also shown to occur for the double occupancies in 
\fig{fig:tUVW_limit}(b), in which $d^2_A$ and $d^2_B$ merge to a unique value at 
$\m=0$. On the other hand, the staggered magnetization $m_z$ remains finite, 
leading to a SDW phase with no charge order. As for the values of the SCO band-gaps, 
which are displayed in \fig{fig:tUVW_limit}(c), they similarly merge towards a unique value upon reduction 
of $\m$ as evidenced by the $(U/t,V/t)=(3.5,0)$ and $(3.5,1)$ plots. 
Moreover, the value of this single gap at $\m=0$ remains $V$-independent, 
which hints at the independence in $V$ of the charge-homogeneous solutions to the SPE discussed below.

In this limit, the model reduces to the $t$--$U$--$V$ model. Its zero temperature 
phase diagram is displayed in \fig{fig:tUV}(a) in dependence on $U$ and $V$. It 
features two phases, namely the SDW and the CDW phases. They are separated by the 
aforementioned $V=U/4$ phase boundary, across which a first order transition occurs. 
Below this phase boundary, the CDW solutions continue to exist in a finite region 
delimited by the PM-CDW instability line (along which $\delta n=0$ and 
$F_{PM}=F_{CDW}$) and the CDW end line (as defined in previous sections). These two lines meet at the 
critical end point (CEP) where the 
PM-CDW instability vanishes. Beyond this point, no CDW solutions with arbitrarily 
small values of $\delta n$ are found. 

\fig{fig:tUV}(c), \fig{fig:tUV}(d), \fig{fig:tUV}(e) and \fig{fig:tUV}(f) 
provide an overview of the variations of $\delta n$ and $d^2$ in the CDW phase 
as functions of $V$ for representative values of $U$: $U=4~t$, $5~t$ and $10~t$, at 
zero temperature. For these values, the CDW-SDW transition occurs at 
$V=t$, $V=1.25~t$ and $V=2.5~t$, respectively. Let us notice that, for large $U$ 
($U>10~t$), the CDW phase closely resembles a pair density wave (PDW), as the 
double occupancy oscillates between nearly zero and nearly one while the 
single occupancy is strongly suppressed. Furthermore, this PDW shows little 
dependence on $V$ when the latter is larger than $U/4$. Upon reducing $U$, 
the PDW gradually turns into a genuine CDW. In all cases, we see that the order 
parameters discontinuously jump across the transition. More generally, no value 
of $U$ has been found to exhibit a continuous CDW-SDW transition. Instead, 
$\delta n$ systematically discontinuously goes from a finite value to zero at 
the CDW-SDW transition point, assessing the phase transition to be of first order. 

\subsection{Finite temperature results}

In \fig{fig:tUV}(b), the phase diagram of the model is shown for $T=t/6$ as a 
function of $U$ and $V$. It exhibits an additional phase at weak 
couplings, namely the PM phase. 
As explained below, the PM-CDW and PM-SDW phase boundaries 
follow from second order phase transitions.
The SDW-CDW transition however remains of 
first order. Similarly to the $T=0$ case, the transition line in the strong 
coupling regime corresponds to the $V=U/4$ line. This result agrees with 
QMC simulations~\cite{zhang1989} and DCA calculations~\cite{terletska2017,
paki2019}. 

In addition, we compared the free energy of the SDW phase at $U=1.9~t$ and $V=0.4~t$ 
and $\beta=100~t/3$ with the determinantal QMC (DQMC) results given in the supplemental 
material of \rref{sushchyev2022}.
They found an energy of $-0.453t$, whereas we find an energy of $-0.299~t$ in the HF 
approximation and $-0.387~t$ in the KRSB formalism. 
Hence, the energy difference between the DQMC and KRSB energies is as small as
$8\times 10^{-2}~W$, with $W$ the bare bandwidth. 
This supports prior agreement between both approaches.
Note that, in the context of the 
t-U Hubbard model, the energy difference between QMC and KRSB is 
largest for U around 2t (see \rref{lilly1990}).

Moreover, the parameter region for the PM phase is in rather good 
quantitative agreement with DCA results. 
We also observed that the PM-CDW transition line deviates from the $V=U/4$ line at weak couplings, 
while a region of CDW-SDW coexistence develops at strong couplings. 
Our PM-SDW phase boundary does not, however, depend on $V$ while DCA predictions point towards a 
weak $V$-dependence. 
We identify two points of particular relevance: the CEP, as defined for \fig{fig:tUV}(a) and 
a triple point (TP) where the PM, CDW and SDW solutions become degenerate. 
Most notably, when compared to \fig{fig:tUV}(b), the CEP is shifted towards weaker couplings.

The only regime in which we can define a continuous transition between the CDW phase and the SDW 
phase is at finite temperature. Since the PM region in \fig{fig:tUV}(b) is bounded 
by the PM-SDW and PM-CDW instability lines, we can take a path in $U$--$V$ space 
starting from, \textit{e.~g.}, the CDW region, going through the PM region and 
finally ending in the SDW region. This would yield a path along which only 
continuous variations of the order parameters are observed, with $\delta n$ 
going from a finite value in the CDW region down to zero in the PM region and 
$m_z$ going from zero in the PM region to a finite value in the SDW region.

It is important to note that, despite close investigation in the region around 
the SDW-CDW transition line and the $F_{PM}=F_{CDW}=F_{SDW}$ triple point, no 
joint spin-and-charge modulated solution to the SPE could be obtained. This, 
and the collapse of the SCO order in the whole $U$--$V$ range in the 
$\m\to 0$ limit, hints that breaking the $n_A=n_B=1$ symmetry of the SDW phase 
in the half-filled $t$--$U$--$V$ model might yield, if any, an order more intricate 
than the SCO one.

\begin{figure}
    \centering
    \includegraphics[width=.48\textwidth]{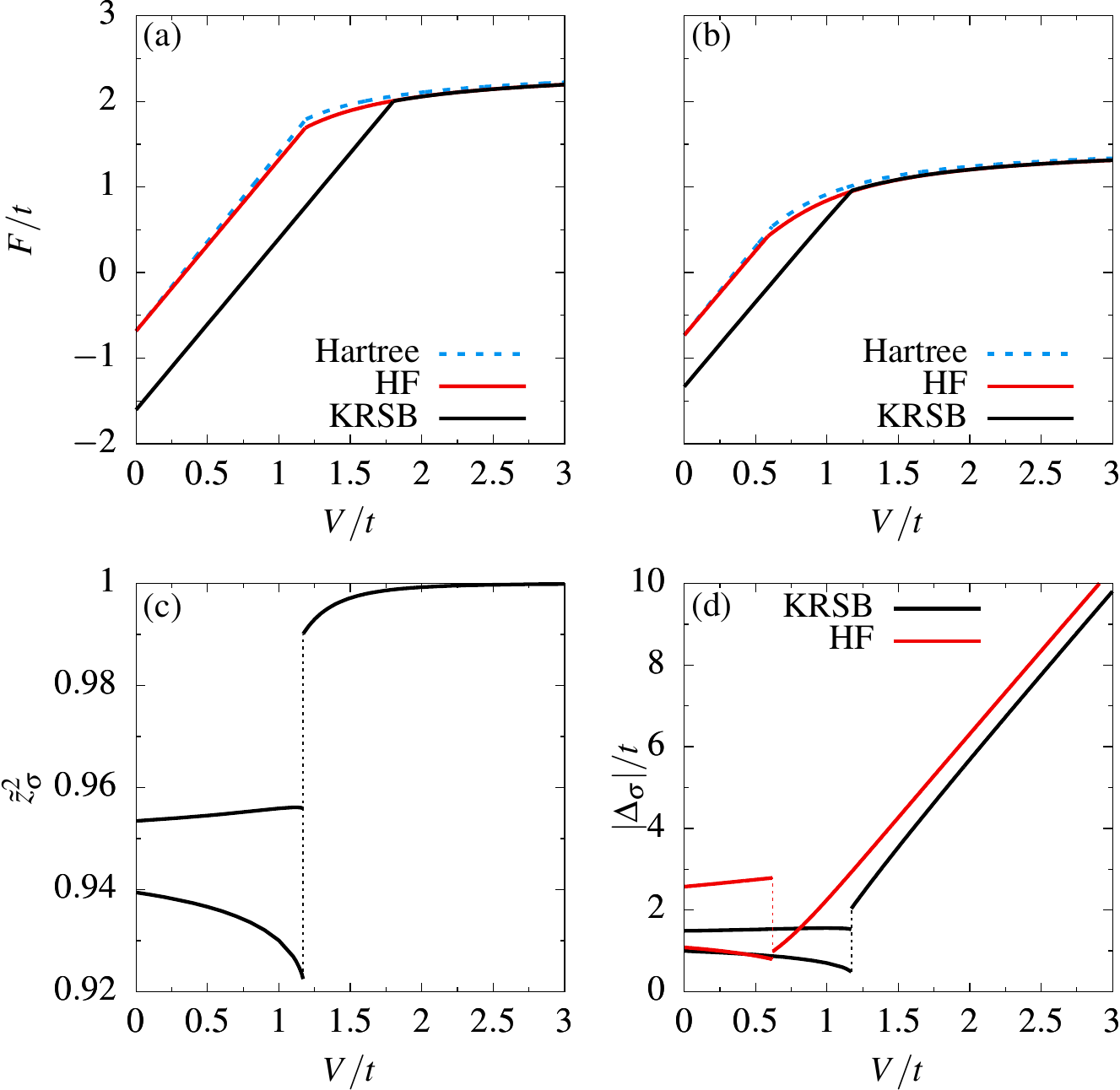}
    \caption{
    Zero temperature free energy obtained from the Hartree (dotted blue), 
    HF (solid red), and KRSB (solid black) calculations as functions of $V$.
    (c) Slave-boson renormalization factors and (d) Slave-boson (black) and HF (red) gaps as
    functions of $V$.
    The dotted lines denote discontinuous jumps at the phase transition.
    Parameters: (a) $U=5~t$, $\m=t/10$, (b), (c) and (d) $U=5~t$, $\m=t$.
    }
    \label{fig:comp}
\end{figure}

\section{Effect of correlations}\label{sec:comp}

In this section, we assess for the relevance of correlations
by comparing the phase diagrams as well as the free energies obtained in the HF approximation 
and in the KRSB formalism.
We also focus on possible markers of a strongly-correlated phase, namely the slave-boson
renormalization factors as well as the difference between the gaps in the dispersion obtained in
each method.

Qualitatively our HF and KRSB calculations support a series of common conclusions.
The phase diagrams are both made of CO and SCO regimes and entail large phase coexistence regions.
Yet, they markedly differ from the quantitative point of view. 
For instance, for $V=t$, the SCO phase may be
found for $U$ as small as $3.4~t$ in KRSB formalism, while it takes $U=4~t$ to stabilize it in the 
HF approximation. In addition, their $(U,\m)$ phase diagram differ in the shape of their phase 
boundaries. Specifically, by continuously increasing $\m$ at fixed $U>3.4~t$, a transition 
from the CO phase to the SCO phase is found, while a second transition back to the CO phase arises 
in the KRSB phase diagram. 
This re-entrant behavior is absent in the HF phase diagram in $(U,\m)$ space where only a single 
phase transition is possible when increasing $\m$.

These discrepancies between the phase diagrams obtained by both methods follow from the 
difference between the free energy obtained in the HF approximation and the one obtained in the 
KRSB representation.
\fig{fig:comp}(a) and \fig{fig:comp}(b) present the zero temperature free energies obtained
in the Hartree approximation, HF approximation and in the KRSB formalism as functions of $V$ for $U=5~t$
and $\m=t/10$ as well as $\m=t$, respectively.
At small values of $V$, the stable phase is the SCO one. In this phase, the leading energy scale is 
that of $U$. This leads to the HF energy being approximately $80\%$ ($58\%$) higher than the KRSB 
energy for $\m=t/10$ ($t$), due to the non-negligible $U$-induced correlation in this phase. 
At larger values of $V$, however, the CO phase is stabilized. In this regime, the dominating 
coupling scale becomes $V$ and the HF and KRSB energies differ by less than $0.1\%$. The HF energy 
is actually slightly lower than its KRSB counterpart. This minor difference can be explained by the
increasing relevance of the exchange term \eq{eq:eff_disp_hf} for larger values of $V$, while no 
such contribution of the exchange to the energy arises in the KRSB formalism.

The slave-boson renormalization factors are displayed in \fig{fig:comp}(c) as functions of $V$, 
for $U=5~t$ and $\m=t$. In the large $V$ regime, we see that $\tilde{z}$ quickly goes to one, in 
which case the effective dispersions \eq{eq:disp_hf} and \eq{eq:disp_sb} seem to closely resemble 
one another. Yet, the differences between the two are sizable as the gap of the HF dispersion is 
larger than its KRSB counterpart. 
As for the SCO phase, we see that, albeit close to one, $\tilde{z}_{\up}$ and $\tilde{z}_{\down}$ 
do not reach one. 
This portrays a slightly correlated phase, the energy of which must then differ from 
the HF energy, as evidenced in the previous paragraph.

In \fig{fig:comp}(d), the band gaps obtained in the HF approximation and in the KRSB representation 
are displayed as functions of $V$, for $U=5~t$ and $\m=t$. We notice that the HF gaps also markedly 
differ from the KRSB gaps, especially in the SCO phase where the largest $|\Delta_\s|$ in the HF 
approximation is approximately $50\%$ larger than its KRSB counterpart. This may also lead to 
free energy discrepancies between both methods.

\section{Conclusion and outlook}\label{sec:concl}

Summarizing, we have set up and addressed a microscopical $t$--$U$--$V$--$\m$
model supporting a spin-and-charge ordered phase in its phase diagram in two
dimensions at half-filling. Such a phase is intuitively expected to smoothly
connect the spin density wave ground state of the half-filled Hubbard model
and the charge density wave of the half-filled $t$--$U$--$V$ extended Hubbard
model. Our findings do not support this expectation and, at zero temperature,
we unraveled discontinuous transitions only, apart from the celebrated
instability to N\'eel order for $U=V=\m=0$. Furthermore, the SCO order
systematically collapses when $\m$ is suppressed down to zero, irrespective of
the values of $U$ and $V$. The splitting is therefore essential to the
stabilization of the SCO phase. 

Let us emphasize that the staggered local potential is also pivotal to the first order phase transition 
that we could relate to resistive switching. As compared to the intensely sought for superconductivity, 
its associated temperature is larger by about one order of magnitude, or even more, making it easier to 
observe.
Beyond materials, cold atoms offer a possibility to experimentally unravel this transition, too, as they can 
easily be submitted to the SLP and since the transition temperature may be of order $t$.

For fixed, moderate $V$, our HF phase diagram of the $t$--$U$--$V$--$\m$ model at half-filling comprises 
both the CO and SCO phases. In the $U$-$\m$ plane, the phase transition is discontinuous and the phase 
boundary is well approximated by a simple straight line $\m_{crit} \propto U$ that is located within a CO 
and SCO coexistence region. In contrast, in the KRSB formalism, the SCO phase displays a re-entrant 
behavior: starting from a critical point ($\m \simeq 0.2~t$, $U \simeq 3.4~t$), two phase boundary lines develop. 
A first one where $\m_{crit}$ increases about linearly with $U$, and a second one where $\m_{crit}$ 
decreases about linearly with $U$. In the former case, the phase boundary corresponds to the end line of 
the SCO phase, while in the latter case we obtain a discontinuous transition inside a CO and SCO 
coexistence region. This robustness of the SCO phase in KRSB formalism is rooted in its ability to jointly 
take magnetism and effective mass renormalization into account, which leads to a sizable lowering of the 
free energy.

There is a lesser degree of divergence when it comes to the comparison of the HF and KRSB phase diagrams 
for fixed $\m$ from the qualitative point of view. Yet, in KRSB, the phase boundary between the SCO and CO 
phases is made of two pieces for both small and large $\m$-values. The first segment, for small 
$V$-values, corresponds to the end points of the SCO phase, while for large $V$-values, it corresponds to 
the free energy crossing of the CO and SCO solutions, in a regime where both phases coexist. 
This transition is discontinuous. In contrast, the HF approach yields this phase boundary to be made 
of the second piece only, at the exception of a small $\m$-small $V$ regime ($V \lesssim 0.2~t$). 
Moreover, no pure SCO phase is predicted at the HF level, while the KRSB approach predicts that only 
the SCO phase is stabilized in the large $U$-smaller $V$ regime ($V < U/4$), irrespective of the value 
of $\m$. That the end line of the CO phase is missed in HF may not be very astonishing, as it happens 
for rather large $U$-values, where HF theory is not controlled.

Regarding the pure $t$--$U$--$V$ model at finite temperature, the ground states
are not systematically ordered any longer, especially in the weak coupling
regime. Accordingly, instability lines towards SDW and CDW phases are found
and it becomes possible to go continuously from the former to the
latter. There is obvious interest to the unraveling of the influence of doping
on the phase diagram, and work towards this end is in progress.


\section*{Acknowledgments}
This work was supported by R\'egion Normandie through the ECOH project. 
Financial support provided by the ANR LISBON (ANR-20-CE05-0022-01) project 
is gratefully acknowledged.
We are thankful to P. Limelette, A. Maignan, D. Pelloquin and M. Raczkowski for fruitful 
discussions.


\appendix

\section{Hartree-Fock self-consistent field equations} \label{app:hf}

Minimizing \eq{model:f_hf} with respect to the different order-parameters yields the following 
equations,
\begin{subequations}
\begin{align}
    &1 = \sum_{\s} \mc{J}_{0,\s} , \\
    &\delta n = \sum_{\s} \mc{J}_{1,\s} , \\
    &m_z = \sum_{\s} \underline{\tau}^3_{\s\s} \mc{J}_{1,\s} , \\
    &b_\up = \mc{K}_{\up} , \\
    &b_\down = \mc{K}_{\down} , \\
    &\delta b_\up = \delta b_\down = 0 ,
\end{align}
\end{subequations}
with
\begin{subequations}
\begin{equation}
    \mc{J}_{m,\s} = \frac{1}{N_L}\sump{\bk,\nu} n_F(E_{\bk,\s,\nu}) \left[ \frac{\nu\Delta^{HF}_{\s}}{\sqrt{(t^{eff}_{\bk,\s})^2+(\Delta^{HF}_{\s})^2}} \right]^m ,
\end{equation}
\begin{equation}
    \mc{K}_{\s} = \frac{1}{N_L}\sump{\bk,\nu} n_F(E_{\bk,\s,\nu}) \frac{\nu t^{eff}_{\bk,\s} (\cos k_x + \cos k_y)}{4\sqrt{(t^{eff}_{\bk,\s})^2+(\Delta^{HF}_{\s})^2}} ,
\end{equation}
\end{subequations}
and where $n_F(E_{\bk,\s,\nu})$ is the Fermi function
\begin{equation}
    n_F(E_{\bk,\s,\nu}) = \frac{1}{1+\exp(\beta E_{\bk,\s,\nu})} .
\end{equation}

\section{Treatment of the slave boson saddle-points} \label{app:saddle_point}

\subsection{Expression of the KRSB grand potential}

We study the saddle-point configurations of the grand potential by imposing symmetries on 
the saddle-point solutions in order to describe the phases that we found to be of 
relevance. They are the paramagnetic (PM), charge ordered (CO), spin density 
wave (SDW), or spin-and-charge order (SCO) phases. None of them carry net 
magnetization. This is taken care of by the following relationship between the 
bosons:
\begin{equation}\label{model:m=0}
    m^{\phantom{X}}_{z,A}+m^{\phantom{X}}_{z,B} = p^2_{A,\up} - p^2_{A,\down} + p^2_{B,\up} - p^2_{B,\down} = 0 .
\end{equation}
It may be satisfied by the relations
\begin{subequations} \label{model:p=q}
\begin{align}
    &p^2_{A,\up}=p^2_{B,\up} , \\
    &p^2_{A,\down}=p^2_{B,\down} ,
\end{align}
\end{subequations}
but this is not the only solution. In fact, the lowest in energy solution  
generally does not fulfill \eq{model:p=q}. Besides, the half filled 
lattice condition is translated as:
\begin{equation}\label{model:n=1}
    n_{A,\s} + n_{B,\s} = p^2_{A,\s} + d^2_A + p^2_{B,\s} + d^2_B = 1 .
\end{equation}
No further symmetry is imposed on an SCO configuration, that corresponds to the lowest symmetry phase. All seven unknown saddle-point parameters for each sublattice may thus be gathered as
\begin{equation} \label{eq:scdw}
    \psi^{SCO}_s = \left( \en_s,\dn_s,0,\pn_{s,\up},\pn_{s,\down},\bn_{s,\up},\bn_{s,\down},\an_s \right) .
\end{equation}
In order to generate a CO configuration, we add another condition, namely that the magnetization vanishes on each sublattice. It reads
\begin{equation}
    m^{\phantom{X}}_{z,s} = p^2_{s,\up} - p^2_{s,\down} = 0 ,
\end{equation}
This yields a charge-ordered nonmagnetic saddle point for which the five unknowns for each sublattice may be gathered as
\begin{equation} \label{eq:cdw}
    \psi^{CO}_s = \left( e_s,d_s,0,\pn_s,\pn_s,\bn_s,\bn_s,\an_s \right) ,
\end{equation}
In the SDW phase, the local magnetization is finite but takes opposite values on each sublattice, so that
\begin{subequations}
\begin{equation}
    m^{\phantom{X}}_{z,A} = -m^{\phantom{X}}_{z,B} \neq 0 .
\end{equation}
Moreover, the density is fixed to one on each sublattice
\begin{equation}
    n_s = p^2_{s,\up} + p^2_{s,\down} + 2d^2_s = 1.
\end{equation}
\end{subequations}
This generates a magnetic saddle point with the six unknowns for each sublattice gathered as
\begin{subequations}
\begin{align} \label{eq:sdw}
    &\psi^{SDW}_A = \left( d,d,0,\pn_\up,\pn_\down,\bn_\up,\bn_\down,\an \right) , \\
    &\psi^{SDW}_B = \left( d,d,0,\pn_\down,\pn_\up,\bn_\down,\bn_\up,\an \right) .
\end{align}
\end{subequations}
Finally, in order to generate the highest symmetry configuration, corresponding to a PM solution, we impose all the above symmetries to the saddle-point solution. This yields a homogeneous mean-field of with a total of only four unknowns:
\begin{equation} \label{eq:pm}
    \psi^{PM}_A = \psi^{PM}_B = \left( d,d,0,\pn,\pn,\bn_0,\bn_0,\an \right) .
\end{equation}

For each of the considered phases, we obtain the fermionic contribution to the
Lagrangian $\mathcal{L}_f$ as
\begin{equation}
\mathcal{L}_f = \sum_{\bk,\s} 
\begin{pmatrix}
\fd_{\bk,\s}, & \fd_{\bk+\bQ,\s}
\end{pmatrix}
\underline{K}^{\phantom{0}}_{\bk,\s} 
\begin{pmatrix}
\fn_{\bk,\s} \\
\fn_{\bk+\bQ,\s}
\end{pmatrix},
\end{equation}
with the pseudofermions inverse propagator
\begin{equation}
\underline{K}^{\phantom{0}}_{\bk,\s} = 
\begin{pmatrix}
\partial_\tau - \mu + \bmoy_{\s} + X_{\bk,\s} 
& \Delta^{SB}_{\s} + Y_{\bk,\s} \\
\Delta^{SB}_{\s} + Y_{\bk,\s} 
& \partial_\tau - \mu + \bmoy_{\s} + X_{\bk+\bQ,\s}
\end{pmatrix},
\end{equation}
with
\begin{align}
    &X_{\bk,\s}  = \zeta^2_{+,\s} \xi_{\bk} + \zeta^2_{-,\s} \xi_{\bk+\bQ} , \\
    &Y_{\bk,\s} = \zeta_{+,\s} \zeta_{-,\s} (\xi_{\bk}+\xi_{\bk+\bQ}) ,
\end{align}
in which the bare dispersion is given by
\begin{equation}
    \xi_{\bk} = \frac{1}{N_L} \sum_{i,j} t^{\phantom{0}}_{ij} \exp \left[-\ii \bk \cdot \left( \bR_i - \bR_j \right) \right] ,
\end{equation}
and where we introduced
\begin{subequations}
\begin{align}
&\bQ=\transp{(\pi,\pi)} ,\\
&\zeta_{\pm,\s} = \frac{1}{2} \left( z_{A,\s} \pm z_{B,\s} \right) .
\end{align}
\end{subequations}
Expressed as such, the inverse propagator may be straightforwardly diagonalized. 
In its eigenmodes basis, we find the following dispersion for the pseudofermions
\begin{align}
    E_{\bk,\s,\nu} &= \bmoy_\s - \mu + \frac{1}{2} \left( z^2_{A,\s} + z^2_{B,\s} \right)  \left( \xi_{\bk} + \xi_{\bk+\bQ} \right) \notag\\
    &\,\, + \nu \sqrt{ \left[ \frac{1}{2} \tilde{z}^2_{\s} \left( \xi_{\bk} - \xi_{\bk+\bQ} \right) \right]^2 + (\Delta^{SB}_{\s})^2} ,
\end{align}
with $\nu=\pm1$. It enters the fermionic contribution to the grand potential in the saddle-point 
approximation which reads
\begin{align}
    \Omega &= \Omega_f+\Omega_b \notag \\
      &= -\frac{1}{\beta} \sump{\bk,\s,\nu} \ln \left[ 1 + \exp \left(-\beta E_{\bk,\s,\nu}\right) \right] \notag \\
      &\; + N_L \Bigg\{ \frac{\an_A}{2} \left( e^2_A + \sum_\s p^2_{A,\s} + \left|d_A\right|^2 - 1 \right) \notag\\ 
      &\;\; + \frac{\an_B}{2} \left( e^2_B + \sum_\s p^2_{B,\s} + \left|d_B\right|^2 - 1 \right) \notag \\
      &\;\; - \sum_\s \left[ \bmoy_\s \avg{n_{\s}} - \frac{\Delta^{SB}_{\s} + \m}{2} ( \delta n + \underline{\tau}^3_{\s\s} m_z ) \right] \notag \\
      &\;\; + \frac{U}{2} \left( \left|d_A\right|^2 + \left|d_B\right|^2 \right) + 2V (\avg{n}^2-\delta n^2) \Bigg\} .
\end{align}

\subsection{Saddle-point equations}

In order to find the saddle-point bosonic configurations, $\psi=\psi_A \oplus \psi_B$, we need to solve 
the system of equations given by $\delta S/\delta\psi=0$. 
This yields a system of fourteen equations (there are seven boson expectation values per 
sublattice), two of which being merely the constraint \eq{model:constraint_a} enforced on average 
on each sublattice. 
We thus need to explicitly solve a system of twelve equations, given by the variations of $\S$ 
with respect to the constraint fields
\begin{subequations}
\begin{align}
    &\I_{0,\up}   = \frac{1}{2} \avg{n_{\up}} , \label{spe:1} \\
    &\I_{0,\down} = \frac{1}{2} \avg{n_{\down}} , \label{spe:2} \\
    &\I_{1,\up}   = \frac{1}{2} ( \delta n + m_z ) , \label{spe:3} \\
    &\I_{1,\down} = \frac{1}{2} ( \delta n - m_z ) , \label{spe:4}
\end{align}
and by the variations of $\S$ with respect to the bosonic fields
\begin{align}
    &\frac{1}{e_A} \sum_\s \E_\s z_{B,\s} \frac{\partial z_{A,\s}}{\partial e_A} = -\alpha_A , \label{spe:5} \\
    &\frac{1}{e_B} \sum_\s \E_\s z_{A,\s} \frac{\partial z_{B,\s}}{\partial e_B} = -\alpha_B , \label{spe:6}
\end{align}
\begin{align}
    &\frac{1}{p_{A,\up}}\! \sum_\s \E_\s z_{B,\s} \frac{\partial z_{A,\s}}{\partial p_{A,\up}} \!=\! - \alpha_A \!+\! \bmoy_\up \!-\! \Delta^{SB}_{\up} \!-\! 4 V ( \avg{n} - \delta n ) \!+\! \m , \label{spe:7} \\
    &\frac{1}{p_{B,\up}}\! \sum_\s \E_\s z_{A,\s} \frac{\partial z_{B,\s}}{\partial p_{B,\up}} \!=\! - \alpha_B \!+\! \bmoy_\up \!+\! \Delta^{SB}_{\up} \!-\! 4 V ( \avg{n} + \delta n ) \!-\! \m , \label{spe:8} \\
    &\frac{1}{p_{A,\down}}\! \sum_\s \E_\s z_{B,\s} \frac{\partial z_{A,\s}}{\partial p_{A,\down}} \!=\! - \alpha_A \!+\! \bmoy_\down \!+\! \Delta^{SB}_{\down} \!-\! 4 V ( \avg{n} - \delta n ) \!+\! \m , \label{spe:9} \\
    &\frac{1}{p_{B,\down}}\! \sum_\s \E_\s z_{A,\s} \frac{\partial z_{B,\s}}{\partial p_{B,\down}} \!=\! - \alpha_B \!+\! \bmoy_\down \!-\! \Delta^{SB}_{\down} \!-\! 4 V ( \avg{n} + \delta n ) \!-\! \m , \label{spe:10}
\end{align}
\begin{equation}
\begin{split}
    \frac{1}{d_A}\! \sum_\s \E_\s z_{B,\s} \frac{\partial z_{A,\s}}{\partial d_A} \!=\! &- \alpha_A \!+\! \sum_{\s} ( \bmoy_\s \!+\! \Delta^{SB}_\s ) \!-\! U \!-\! 8 V ( \avg{n} - \delta n ) \\
    &\!+\! 2\m , \label{spe:11} \\
\end{split}
\end{equation}
\begin{equation}
\begin{split}
    \frac{1}{d_B}\! \sum_\s \E_\s z_{A,\s} \frac{\partial z_{B,\s}}{\partial d_B} \!=\! 
    &- \alpha_B \!+\! \sum_{\s} ( \bmoy_\s \!-\! \Delta^{SB}_\s ) \!-\! U \!-\! 8 V ( \avg{n} + \delta n ) \\
    &\!-\! 2\m . \label{spe:12}
\end{split}
\end{equation}
\end{subequations}\label{eq:spe}
Here, we introduced the shorthand notations
\begin{subequations}
\begin{equation}
    \I_{m,\s} = \sump{\bk,\nu} n_F \left( E_{\bk,\s,\nu} \right) \left[ \frac{\nu \Delta^{SB}_{\s}}{\sqrt{(\tilde{z}^2_{\s}t_{\bk})^2+(\Delta^{SB}_{\s})^2}} \right]^m ,
\end{equation}
and
\begin{equation}
    \E_\s = \sump{\bk,\nu} n_F \left( E_{\bk,\s,\nu} \right) 
    \frac{\nu \tilde{z}^2_{\s}t^2_{\bk}}{\sqrt{(\tilde{z}^2_{\s}t_{\bk})^2+(\Delta^{SB}_{\s})^2}} .
\end{equation}
\end{subequations}

In the CO phase, this number of unknowns --- and hence the dimension of the corresponding system 
of equations --- is reduced by four. Indeed, it is straightforward to verify that \eq{spe:1} and 
\eq{spe:2} become equivalent, so do \eq{spe:3} and \eq{spe:4}, \eq{spe:7} and \eq{spe:9} as well 
as \eq{spe:8} and \eq{spe:10}. We are thus left with a system of equations of dimension eight. 
In the $\m=0$ case, the number of unknowns is further reduced by three, since \eq{spe:7} and 
\eq{spe:8} (hence \eq{spe:9} and \eq{spe:10}) become equivalent, implying that \eq{spe:5} and 
\eq{spe:12} as well as \eq{spe:6} and \eq{spe:11} also become equivalent. 
This leads to a system of equations of dimension five.

In the SDW phase, \eq{spe:1} and \eq{spe:2} become equivalent, so do \eq{spe:3} and \eq{spe:4}, 
\eq{spe:7} and \eq{spe:10}, \eq{spe:8} and \eq{spe:9}, as well as \eq{spe:5}, \eq{spe:6}, 
\eq{spe:11} and \eq{spe:12}. This reduces the number of unknowns by seven, leading to a system of 
equations of dimension five.

Finally, in the PM phase both sublattices are strictly equivalent and both spin projections 
$\s=\up,\down$ are equivalent. It has been shown that in this specific case, the saddle-point 
equations reduce to a single equation \cite{kotliar1986}
\begin{equation}
    1-4d^2=-\frac{U}{8 \epsilon_0},
\end{equation}
with 
\begin{align}
    \epsilon_0=\frac{2}{N_L} \sum_{\bk} n_F \left( E_{\bk} \right) t_{\bk},
\end{align}
reproducing the seminal result from Brinkman and Rice \cite{brinkman1970}.

%

\end{document}